\documentclass[12pt]{JHEP3}

\usepackage{epsfig}
\usepackage{subfigure}
\usepackage{wrapfig}

\usepackage{amsmath}
\usepackage{amssymb}
\usepackage{cite}
\usepackage{array}

\newcommand{\arcsinh}{{\rm arcsinh}\,}
\newcommand{\sign}{{\rm sign}\,}

\newcommand{\beqs}{\begin{equation*}}
\newcommand{\beq}{\begin{equation}}

\newcommand{\eeqs}{\end{equation*}}
\newcommand{\eeq}{\end{equation}}

\newcommand{\beqas}{\begin{eqnarray*}}
\newcommand{\beqa}{\begin{eqnarray}}

\newcommand{\eeqas}{\end{eqnarray*}}
\newcommand{\eeqa}{\end{eqnarray}}

\newcommand{\eq}[2]{\begin{equation} #1 \label{#2} \end{equation}}

\newcommand{\eps}{\varepsilon}
\newcommand{\al}{\alpha}
\newcommand{\be}{\beta}
\newcommand{\ga}{\gamma}
\newcommand{\de}{\delta}
\newcommand{\om}{\omega}
\newcommand{\ka}{\kappa}
\newcommand{\la}{\lambda}

\newcommand{\Om}{\Omega}

\newcommand{\La}{\Lambda}

\newcommand{\blist}{\begin{itemize}}

\newcommand{\elist}{\end{itemize}}

\providecommand{\href}[2]{#2}

\DeclareFontFamily{OT1}{rsfs}{}
\DeclareFontShape{OT1}{rsfs}{m}{n}{ <-7> rsfs5 <7-10> rsfs7 <10->rsfs10}{} 
\DeclareMathAlphabet{\mycal}{OT1}{rsfs}{m}{n}

\newcommand{\plr}[1]{\overleftrightarrow{\partial_{#1}}}

\newcommand{\chib}{\overline{\chi}}
\newcommand{\chid}{\chi^*}

\usepackage{slashed}

\newcommand{\plrd}[1]{\plr{#1}}

\def\cL{{\cal L}}

\newcommand{\diff}{\extd}

\newcommand{\act}{S_{\rm 2DG}}

\DeclareMathOperator{\extdm}{d} 
\newcommand{\extd}{\extdm \!}

\newcommand{\tac}{{\mathcal T}}

\title{Ramifications of Lineland}

\author{Daniel Grumiller\footnotemark[1]\,\, and Rene Meyer\footnotemark[1]\,\,\footnotemark[2]\\ 

\footnotemark[1]\,\,\parbox[t]{13cm}{Institute for Theoretical Physics, University of Leipzig\\ Augustusplatz 10-11, D-04109 Leipzig, Germany} \\ \ \\

\footnotemark[2]\,\,\parbox[t]{13cm}{Max Planck Institute for
Mathematics in the Sciences\\ Inselstrasse 22, D-04103 Leipzig, Germany} \\ \ \\

E-mail: \email{grumiller@itp.uni-leipzig.de}, \email{Rene.Meyer@itp.uni-leipzig.de}.}

\abstract{ A non-technical overview on gravity in two dimensions is
  provided. Applications discussed in this work comprise 2D type 0A/0B
  string theory, Black Hole evaporation/thermodynamics, toy models for
  quantum gravity, for numerical General Relativity in the context of
  critical collapse and for solid state analogues of Black Holes.
  Mathematical relations to integrable models, non-linear gauge
  theories, Poisson-sigma models, KdV surfaces and non-commutative
  geometry are presented.  }


\preprint{LU-ITP-2006/004}

\begin{document}


\section{Introduction} 

The study of gravity in 2D --- boring to some, fascinating to others
\cite{Square:1884} --- has the undeniable disadvantage of eliminating
a lot of structure that is present in higher dimensions; for instance,
the Riemann tensor is determined already by the Ricci scalar, i.e.,
there is no Weyl curvature and no trace-free Ricci part. On the other
hand, it has the undeniable advantage of eliminating a lot of
structure that is present in higher dimensions; for instance,
non-perturbative results may be obtained with relative ease due to
technical simplifications, thus allowing one to understand some
important conceptual issues arising in classical and quantum gravity
which are universal and hence of relevance also for higher
dimensions. 

The scope of this non-technical overview is broad rather than
focussed, since there exist already various excellent reviews and
textbooks presenting the technical pre-requisites in
detail,\footnote{For instance, the status of the field in the late
  1980ies is summarized in \cite{Brown:1988}.} and because the
broadness envisaged here may lead to a cross-fertilization between
otherwise only loosely connected communities.
Some recent results are presented in more detail. 
It goes without saying that the topics selected concur with the
authors' preferences; by no means it should be concluded that an issue
or a reference omitted here is devoid of interest.

The common link between all applications mentioned here is 2D dilaton gravity,%
\footnote{The 2D Einstein-Hilbert action will not be discussed except in section \ref{se:EH2D}.}
\begin{equation}
\label{eq:GDT}
\act=\frac12 \int \extd^{2}x\, \sqrt{-g}\; \left[ X R+ U(X)\; (\nabla X)^{2} - 2V(X)\; \right] \, ,
\end{equation}
the action of which depends functionally on the metric $g_{\mu\nu}$
and on the scalar field $X$. Note that very often, in particular in
the context of string theory, the field redefinition $X=e^{-2\phi}$ is
employed; the field $\phi$ is the dilaton of string theory, hence the
name ``dilaton gravity''. However, it is emphasized that the natural
interpretation of $X$ need not be the one of a dilaton field --- it
may also play the role of surface area, dual field strength,
coordinate of a suitable target space or black hole (BH) entropy,
depending on the application.  The curvature scalar $R$ and covariant
derivative $\nabla$ are associated with the Levi-Civita connection and
Minkowskian signature is implied unless stated otherwise. The
potentials $U$, $V$ define the model; several examples will be
provided below. A summary is contained in table \ref{tab:1}.

This proceedings contribution is organized as follows: section
\ref{se:1} is devoted to a reformulation of \eqref{eq:GDT} as a
non-linear gauge theory, which considerably simplifies the
construction of all classical solutions; section \ref{se:2} discusses
applications in 2D string theory; section \ref{se:3} summarizes
applications in BH
physics; 
section \ref{se:4} demonstrates how to reconstruct geometry from
matter in a quantum approach; section \ref{se:5} contains not only
mathematical issues but also some open problems.

\section{Gravity as non-linear gauge theory}\label{se:1}

It has been known for a long time how to obtain all classical
solutions of \eqref{eq:GDT} not only locally, but globally. Two
ingredients turned out to be extremely useful: a reformulation of
\eqref{eq:GDT} as a first order action and the imposition of a
convenient (axial or Eddington-Finkelstein type) gauge, rather than
using conformal gauge.\footnote{In string theory almost exclusively
  conformal gauge is used. A notable exception is
  \cite{Polyakov:1987zb}.}  Subsequently we will briefly recall these
methods. For a more comprehensive review cf.~\cite{Grumiller:2002nm}.

\begin{table}
\centering\hspace*{-0.8truecm}
\fbox{
\begin{tabular}{|l||>{$}c<{$}|>{$}c<{$}||>{$}c<{$}|} 
\hline
Model (cf.~\eqref{eq:GDT} or \eqref{eq:FOG})& U(X) & V(X) & w(X) $\,\,(cf.~\eqref{eq:wI})$ \\ \hline \hline
1.~Schwarzschild  \cite{Thomi:1984na} 
& -\frac{1}{2X} & -\lambda^2 & -2\la^2\sqrt{X} \\
2.~Jackiw-Teitelboim \cite{Teitelboim:1983ux,Jackiw:1985je} & 0 & -\Lambda X & -\frac12 \Lambda X^2 \\ 
3.~Witten BH/CGHS \cite{Witten:1991yr,Callan:1992rs} & -\frac{1}{X} & -2b^2 X & -2b^2X \\
4.~CT Witten BH \cite{Witten:1991yr,Callan:1992rs} & 0 & -2b^2  & -2b^2X \\
5.~SRG ($D>3$) & -\frac{D-3}{(D-2)X} & -\lambda^2 X^{(D-4)/(D-2)} & -\la^2\frac{D-2}{D-3} X^{(D-3)/(D-2)}\\  
6.~$(A)dS_2$ ground state \cite{Lemos:1994py} &  -\frac{a}{X} & -\frac{B}{2}X  & a\neq 2:\,\,-\frac{B}{2(2-a)} X^{2-a}\\
7.~Rindler ground state \cite{Fabbri:1996bz} & -\frac{a}{X} & -\frac{B}{2} X^a  & -\frac{B}{2} X \\
8.~BH attractor \cite{Grumiller:2003hq} & 0 & -\frac{B}{2}X^{-1} & -\frac{B}{2}\ln{X} \\ \hline
9.~All above: $ab$-family \cite{Katanaev:1997ni} & -\frac{a}{X} & -\frac{B}{2} X^{a+b} & b\neq-1:\,\,-\frac{B}{2(b+1)}X^{b+1} \\   \hline 
10.~Liouville gravity \cite{Nakayama:2004vk} & a & b e^{\al X} & a\neq-\al:\,\,\frac{b}{a+\al}e^{(a+\al)X} \\
11.~Scattering trivial \cite{Grumiller:2002dm} & $generic$  & 0 & $const.$ \\
12.~Reissner-Nordstr\"om \cite{Reissner:1916} & -\frac{1}{2X} & -\lambda^2 + \frac{Q^2}{X} & -2\la^2\sqrt{X}-2Q^2/\sqrt{X}\\
13.~Schwarzschild-$(A)dS$ \cite{Hawking:1982dh} & -\frac{1}{2X} & -\lambda^2 - \ell X & -2\la^2\sqrt{X} - \frac23 \ell X^{3/2} \\
14.~Katanaev-Volovich \cite{Katanaev:1986wk} & \alpha & \beta X^2 - \Lambda  & \int^X e^{\al y}(\be y^2-\La)\extd y\\
15.~Achucarro-Ortiz \cite{Achucarro:1993fd} & 0 &  \frac{Q^2}{X} - \frac{J}{4X^3} - \Lambda X &  Q^2\ln{X} + \frac{J}{8X^2} - \frac12 \La X^2 \\
16.~KK reduced CS \cite{Guralnik:2003we,Grumiller:2003ad} & 0 & \frac12 X(c-X^2) & -\frac18 (c-X^2)^2 \\ 
17.~Symmetric kink \cite{Bergamin:2005au} & {\rm generic} & -X\Pi_{i=1}^n(X^2-X_i^2) & $cf.~\cite{Bergamin:2005au}$ \\
18.~2D type 0A/0B \cite{Douglas:2003up,Gukov:2003yp} & -\frac{1}{X} & -2b^2X+\frac{b^2q^2}{8\pi}  & -2b^2X+\frac{b^2q^2}{8\pi}\ln{X}\\
19.~exact string BH \cite{Dijkgraaf:1992ba,Grumiller:2005sq} & $\eqref{eq:solutionofESBH3}$ & $\eqref{eq:solutionofESBH3}$  & $\eqref{eq:wESBHNS}$ \\
\hline
\end{tabular}
}
\caption{Selected list of models}
\label{tab:1}
\end{table}

\subsection{First order formulation}\label{se:FOG}

The Jackiw-Teitelboim model (cf.~the second model in table
\eqref{tab:1}) allows a gauge theoretic formulation based upon
$(A)dS_2$, 
\begin{equation}
  \label{eq:ads2}
  [P_a,P_b]=\La\eps_{ab} J\,,\qquad [P_a,J]=\eps_a{}^b P_b\,,
\end{equation}
with Lorentz generator $J$, translation generators $P_a$ and $\La\neq 0$.
A corresponding first order action, $S=\int X_A F^A$, has been
introduced in \cite{Isler:1989hq}. The field
strength $F=\extd A + [A,A]/2$ contains the $SO(1,2)$ connection
$A=e^aP_a+\om J$, and the Lagrange multipliers $X_A$ transform under
the coadjoint representation. This example is exceptional insofar as
it allows a formulation in terms of a {\em linear} (Yang-Mills type)
gauge theory. Similarly, the fourth model in table \ref{tab:1} allows
a gauge theoretic formulation \cite{Verlinde:1991rf} based upon the
centrally extended Poincar{\`e} algebra
\cite{Cangemi:1992bj}. The generalization to
non-linear gauge theories \cite{Ikeda:1993aj} allowed a comprehensive
treatment of all models \eqref{eq:GDT} with $U=0$, which has been
further generalized to $U\neq 0$ in \cite{Schaller:1994es}. The
corresponding first order gravity action%
%
%
\eq{
S_{\rm FOG}= - \int  \left[X_aT^a+XR+\epsilon\left(X^+X^- U(X) + V(X)\right)\right]
}{eq:FOG}
is equivalent to \eqref{eq:GDT} (with the same potentials $U,V$) upon
elimination of the auxiliary fields $X_a$ and the torsion-dependent
part of the spin-connection. 
Here is our notation: $e^a=e^a_\mu dx^\mu$ is the dyad
1-form. 
Latin indices refer to an anholonomic frame, Greek indices to a
holonomic one. The 1-form $\omega$ represents the spin-connection
$\om^a{}_b=\eps^a{}_b\om = \eps^a{}_b\om_\mu\extd x^\mu$ with the
totally antisymmetric Levi-Civita symbol $\eps_{ab}$ ($\eps_{01}=+1$).
With the flat metric $\eta_{ab}$ in light-cone coordinates
($\eta_{+-}=1=\eta_{-+}$, $\eta_{++}=0=\eta_{--}$) it reads
$\eps^\pm{}_\pm=\pm 1$. The torsion 2-form present in the first term
of \eqref{eq:FOG} is given by $T^\pm=(\extd\pm\omega)\wedge e^\pm$.
The curvature 2-form $R^a{}_b$ can be represented by the 2-form $R$
defined by $R^a{}_b=\eps^a{}_b R$ with $R=\extd\om$. It appears in the
second term in \eqref{eq:FOG}.  Since no confusion between 0-forms and
2-forms should arise the Ricci scalar is also denoted by $R$.  The
volume 2-form is denoted by $\epsilon = e^+\wedge e^-$. Signs and
factors of the Hodge-$\ast$ operation are defined by $\ast\epsilon=1$.
It should be noted that \eqref{eq:FOG} is a specific Poisson-sigma
model \cite{Schaller:1994es} with a 3D target space, with target space
coordinates $X,X^\pm$, see section \ref{se:psm} below.
A second order action similar to
\eqref{eq:GDT} has been introduced in \cite{Russo:1992yg}.

\subsection{Generic classical solutions}\label{se:sol}

It is useful to introduce the following combinations of the potentials
$U$ and $V$:
\begin{equation}
  \label{eq:wI}
  I(X):=\exp{\int^X U(y)\extd y}\,,\quad w(X):=\int^XI(y)V(y)\extd y
\end{equation}
The integration constants may be absorbed, respectively, by rescalings
and shifts of the ``mass'', see equation \eqref{eq:c} below. Under
dilaton dependent conformal transformations $X^a\to X^a/\Om$, $e^a\to
e^a\Om$, $\om\to\om+X_ae^a\extd\,\ln{\Om}/\extd X$ the action
\eqref{eq:FOG} is mapped to a new one of the same type with
transformed potentials $\tilde{U}$, $\tilde{V}$. Hence, it is not
invariant. It turns out that only the combination $w(X)$ as defined in
\eqref{eq:wI} remains invariant, so conformally invariant quantities
may depend on $w$ only. Note that $I$ is positive apart from eventual
boundaries (typically, $I$ may vanish in the asymptotic region and/or
at singularities). One may transform to a conformal frame with $\tilde{I}=1$,
solve all equations of motion and then perform the inverse
transformation. Thus, it is sufficient to solve the classical
equations of motion for $\tilde{U}=0$,
\begin{gather}
  \extd X + \tilde{X}^-\tilde{e}^+ - \tilde{X}^+\tilde{e}^- = 0 \label{eq:gPSMeom3.1}\ , \\
  (\extd\pm\tilde{\om}) \tilde{X}^\pm \mp \tilde{e}^\pm \tilde{V}(X) = 0 \label{eq:gPSMeom3.2}\ ,\\
  (\extd\pm\tilde{\om}) \wedge \tilde{e}^\pm = 0 \label{eq:gPSMeom4.2}\ , 
\end{gather}
which is what we are going to do now. Note that the equation
containing $\extd\tilde{\om}$ is redundant, whence it is not
displayed.

Let us start with an assumption: $\tilde{X}^{+}\neq 0$ for a given
patch.
  To get some physical intuition as to what this
  condition could mean: the quantities $X^a$, which are the Lagrange
  multipliers for torsion, can be expressed as directional derivatives
  of the dilaton field by virtue of \eqref{eq:gPSMeom3.1} (e.g.~in the
  second order formulation a term of the form $X^aX_a$ corresponds to
  $(\nabla X)^2$). For those who are familiar with the Newman-Penrose
  formalism: for spherically reduced gravity the quantities $X^a$
  correspond to the expansion spin coefficients $\rho$ and
  $\rho^\prime$ (both are real). 
If $\tilde{X}^{+}$ vanishes a (Killing) horizon
is encountered and one can repeat the calculation below with indices
$+$ and $-$ swapped everywhere. If both vanish in an open region by
virtue of \eqref{eq:gPSMeom3.1} a constant dilaton vacuum emerges,
which will be addressed separately below. If both vanish on isolated
points the Killing horizon bifurcates there and a more elaborate
discussion is needed \cite{Klosch:1996qv}. The patch implied by
$\tilde{X}^+\neq 0$ is a ``basic Eddington-Finkelstein patch'', i.e.,
a patch with a conformal diagram which, roughly speaking, extends over
half of the bifurcate Killing horizon and exhibits a coordinate
singularity on the other half.  In such a patch one may redefine
$\tilde{e}^{+}=\tilde{X}^{+} Z$ with a new 1-form $Z$. Then
\eqref{eq:gPSMeom3.1} implies $\tilde{e}^{-}=\extd
X/\tilde{X}^{+}+\tilde{X}^{-}Z$ and the volume form reads
$\tilde{\epsilon}=\tilde{e}^{+}\wedge \tilde{e}^{-}=Z\wedge \extd X$.
The $+$ component of \eqref{eq:gPSMeom3.2} yields for the connection
$\tilde{\omega}=-\extd \tilde{X}^{+}/\tilde{X}^{+}+Z\tilde{V}(X)$. One
of the torsion conditions \eqref{eq:gPSMeom4.2} then leads to $\extd
Z=0$, i.e., $Z$ is closed. Locally (in fact, in the whole patch) it is
also exact: $Z=\extd u$. It is emphasized that, besides the
integration of \eqref{eq:conservation} below, this is the only
integration needed! After these elementary steps one obtains already
the conformally transformed line element in Eddington-Finkelstein (EF)
gauge
\begin{equation} 
  \label{eq:lieelement}
  \extd \tilde{s}^2=2\tilde{e}^{+}\tilde{e}^{-}=2\extd u\,\extd X + 2\tilde{X}^{+}\tilde{X}^{-}\extd u^2\,,
\end{equation}
which nicely demonstrates the power of the first order formalism. In
the final step the combination $\tilde{X}^+\tilde{X}^-$ has to be
expressed as a function of $X$. This is possible by noting that the
linear combination $\tilde{X}^+\times$[\eqref{eq:gPSMeom3.2} with $-$
index] + $\tilde{X}^-\times$[\eqref{eq:gPSMeom3.2} with $+$ index]
together with \eqref{eq:gPSMeom3.1} establishes a conservation
equation,
\begin{equation}
  \label{eq:conservation}
  \extd{(\tilde{X}^+\tilde{X}^-)} + \tilde{V}(X) \extd X = \extd{(\tilde{X}^+\tilde{X}^- + w(X))} = 0\,.
\end{equation}
Thus, there is always a conserved quantity ($\extd M=0$), which in the
original conformal frame reads
\begin{equation}
  \label{eq:c}
  M=-X^+X^-I(X)-w(X)\,,
\end{equation}
where the definitions \eqref{eq:wI} have been inserted. It should be
noted that the two free integration constants inherent to the
definitions \eqref{eq:wI} may be absorbed by rescalings and shifts of
$M$,
respectively. 
The classical solutions are labelled by $M$, which may be interpreted
as mass (see section \ref{se:td}). Finally, one has to transform back
to the original conformal frame (with conformal factor $\Om=I(X)$).
The line element \eqref{eq:lieelement} by virtue of \eqref{eq:c} may
be written as
\begin{equation}
  \label{eq:EF}
  \extd s^2 =2I(X)\diff u\,\diff X - 2I(X)(w(X) + M)\diff u^2\,. 
\end{equation}
Evidently there is always a Killing vector
$K\cdot\partial=\partial/\partial u$ with associated Killing norm
$K^2=-2I(w+M)$. Since $I\neq 0$ Killing horizons are encountered at
$X=X_h$ where $X_h$ is a solution of
\begin{equation}
  \label{eq:horizon}
  w(X_h) + M =0\,.
\end{equation} 
It is recalled that \eqref{eq:EF} is valid in a basic EF patch, e.g.,
an outgoing one. By redoing the derivation above, but starting from
the assumption $X^-\neq 0$ one may obtain an ingoing EF patch, and by
gluing together these patches appropriately one may construct the
Carter-Penrose diagram,
cf.~\cite{Klosch:1996fi,Klosch:1996qv,Grumiller:2002nm}.

As pointed out in the introduction the full geometric information
resides in the Ricci scalar. The one related to the generic solution
\eqref{eq:EF} reads
\begin{equation}
  \label{eq:Ricci}
  R=\frac{2}{I(X)} \frac{\extd}{\extd X}\Big(U(X)(M+w(X)) + I(X)V(X)\Big) \,. 
\end{equation}
There are two important special cases: for $U=0$ the Ricci scalar
simplifies to $R=2V'(X)$, while for $w(X) \propto 1/I(X)$ it scales
proportional to the mass, $R=2M U'(X)/I(X)$. The latter case comprises
so-called Minkowskian ground state models (for examples cf.~the first,
third, fifth and last line in table \ref{tab:1}). Note that for many
models in table \ref{tab:1} the potential $U(X)$ has a singularity at
$X=0$ and consequently a curvature singularity arises.

\subsection{Constant dilaton vacua}

For sake of completeness it should be mentioned that in addition to
the family of generic solutions \eqref{eq:EF}, labelled by the mass
$M$, isolated solutions may exist, so-called constant dilaton vacua
(cf.~e.g.~\cite{Bergamin:2005au}), which have to
obey\footnote{Incidentally, for the generic case \eqref{eq:EF} the
  value of the dilaton on an extremal Killing horizon is also subject
  to these two constraints.} $X=X_{\rm CDV}=\rm const.$ with
$V(X_{CDV}) =
0$. 
The corresponding geometry has constant curvature, i.e., only
Minkowski, Rindler or $(A)dS_2$ are possible space-times for constant
dilaton vacua.\footnote{In quintessence cosmology in 4D such solutions
  serve as late time $dS_4$ attractor \cite{Hao:2003aa}. In 2D dilaton
  supergravity solutions preserving both supersymmetries are
  necessarily constant dilaton vacua \cite{Bergamin:2003mh}.}  The
Ricci scalar is determined by
\begin{equation}
  \label{eq:CDVR}
  R_{\rm CDV} = 2V'(X_{\rm CDV}) = \rm const. 
\end{equation}
 Examples are provided by the last eighth entries in table \ref{tab:1}. 
For instance, 2D type 0A strings with an equal number $q$ of electric and magnetic D0 branes (cf.~the penultimate entry in table \ref{tab:1}) allow for an $AdS_2$ vacuum with $X_{\rm CDV}=q^2/(16\pi)$ and $R_{\rm CDV}=-4b^2$ \cite{Thompson:2003fz}.

\subsection{Topological generalizations}\label{se:top}

In 2D there are neither gravitons nor photons, i.e.~no propagating physical
modes exist \cite{Birmingham:1991ty}. This feature 
makes the inclusion of Yang-Mills fields in 2D dilaton gravity or an
extension to supergravity straightforward. 
Indeed, both generalizations can be treated again in the first order formulation as a Poisson-sigma model, cf.~e.g.~\cite{Strobl:1999wv}. In addition to $M$ (see \eqref{eq:c}) more locally 
conserved quantities (Casimir functions) may emerge and the 
integrability concept is extended.

As a simple example we include an abelian Maxwell field, i.e., instead of \eqref{eq:FOG} we take  
\begin{equation}
  \label{eq:fogu1}
  S_{\rm MDG}= - \int \left[X_aT^a+XR+BF+\epsilon\left(X^+X^-U(X,B)+V(X,B)\right)\right]\,,
\end{equation}
where $B$ is an additional scalar field and $F=\extd A$ is the field
strength 2-form. Variation with respect to $A$ immediately establishes
a constant of motion, $B=Q$, where $Q$ is some real constant, the
$U(1)$ charge. Variation with respect to $B$ may produce a relation
that allows to express $B$ as a function of the dilaton and the dual
field strength $\ast F$. For example, suppose that
$V(X,B)=V(X)+\frac12 B^2$.  Then, variation with respect to $B$ gives
$B=-{\ast F}$. Inserting this back into the action yields a standard
Maxwell term.
The solution of the remaining equations of motion reduces to the case
without Maxwell field. One just has to replace $B$ by its on-shell
value $Q$ in the potentials $U$, $V$.

Concerning supergravity we just mention a couple of references for
further orientation
\cite{Park:1993sd,Bergamin:2002ju,Bergamin:2003mh}.

\subsection{Non-topological generalizations}

To get a non-topological theory one can add scalar or fermionic
matter. The action for a real, self-interacting and non-minimally
coupled scalar field $\tac$,
\begin{equation}
  \label{eq:cc2}
  S_{\tac}=\frac12 \int \Big[ F(X)\extd\tac\wedge{\ast\extd\tac} + \epsilon f(X,\tac)\Big]\, , 
\end{equation}
in our convention requires $F<0$ for the kinetic term to have the
correct sign; e.g.~$F=-\ka$ or $F=-\ka X$.

While scalar matter couples to the metric and the dilaton, fermions%
\footnote{We use the same definition for the Dirac matrices as in
  \cite{Meyer:2005fz}.} couple directly to the Zweibein
($A\overleftrightarrow{\extd}B = A \extd B - (\extd A)B$),
\begin{equation}
  \label{eq:cc2ferm}
  S_{\chi} =  \int \Big[\frac{i}{2} F(X)\;(*e^a) \wedge (\chib \ga_a \overleftrightarrow{\extd}\chi) + \epsilon H(X) g(\chib\chi)\Big]\, ,
\end{equation}
but not --- and this is a peculiar feature of 2D --- to the spin
connection. The self-interaction is at most quartic (a constant term may be absorbed in $V(X)$),
\eq{g(\chib\chi) = m \chib\chi + \la (\chib\chi)^2\,.}{eq:fermSI}
The quartic term (henceforth: Thirring term \cite{Thirring:1958in})
can also be recast into a classically equivalent form by introducing
an auxiliary vector potential,
\eq{\la \int \epsilon (\chib\chi)^2 = \frac{\la}{2} \int  \left[A\wedge \ast A + 2 A \wedge (\ast e_a) \chib \ga^a \chi\right] \,,}{eq:rewritethirring}
which lacks a kinetic term and thus does not
propagate by itself. 

We speak of minimal coupling if the coupling functions
$F(X),f(X,\tac),H(X)$ do not depend on the dilaton $X$, and of
nonminimal coupling otherwise.

As an illustration we present the spherically reduced
Einstein-massless-Klein-Gordon model (EMKG). It emerges from
dimensional reduction of 4D Einstein-Hilbert (EH) gravity (cf.~the
first model in table \ref{tab:1}) with a minimally coupled scalar
field, with the choices $f(X,\tau)=0$ and
\begin{equation}
  \label{eq:cc22}
  w(X)=-2\la^2\sqrt{X}\,,\quad F(X)=-\ka X\,,\quad I(X)=\frac{1}{\sqrt{X}} \,,
\end{equation}
where $\la$ is an irrelevant scale parameter and $\ka$ encodes the
(also irrelevant) Newton coupling.  Minimally coupled Dirac fermions
in four dimensions yield upon dimensional reduction two 2-spinors
coupled to each other through intertwinor terms, which is not covered
by \eqref{eq:cc2ferm} (see \cite{Balasin:2004gf} for details on
spherical reduction of fields of arbitrary spin and the spherical
reduced standard model).

With matter the equation of motion \eqref{eq:gPSMeom3.2} and the
conservation law \eqref{eq:conservation} obtain contributions $W^\pm =
\de (S_\tac+S_\chi)/\de e^\mp$ and $X^-W^+ + X^+W^-$, respectively,
destroying integrability because $Z$ is not closed anymore: $\extd Z =
W^+\wedge Z/X^+$.  In special cases exact solutions can be obtained:
\begin{enumerate}
\item For (anti-)chiral fermions and (anti-)selfdual scalars with
  $W^+=0$ ($W^-=0$) the geometric solution \eqref{eq:lieelement} is still valid
  \cite{Grumiller:2002nm} and the second equation of motion \eqref{eq:gPSMeom3.2}
  implies $W^- = W^-_u \extd u$. Such solutions have been studied e.g.~in \cite{Kummer:1992ef,Pelzer:1998ea}. They arise also in the Aichelburg-Sexl limit \cite{Aichelburg:1971dh} of boosted BHs \cite{Balasin:2003cn}. 
\item A one parameter family of static solutions of the EMKG 
  has been discovered in \cite{Fisher:1948yn}. Studies of static
  solutions in generic dilaton gravity may be found in
  \cite{Filippov:2002sp,Grumiller:2004wi}. A static solution for the
  line-element with time-dependent scalar field (linear in time) has
  been discussed for the first time in \cite{Wyman:1981bd}. It has
  been studied recently in more detail in \cite{Bilge:2005vx}.
\item A (continuously) self-similar solution of the EMKG has been
  discoverd in \cite{Roberts:1989sk}.
\item Specific models allow for exact solutions even in the presence
  of more general matter sources; for instance, the conformally
  transformed CGHS model (fourth in table \ref{tab:1}), Rindler ground
  state models (seventh in table \ref{tab:1}) and scattering trivial
  models (eleventh in table \ref{tab:1}).
\end{enumerate}


\section{Strings in 2D} \label{se:2}

Strings propagating in a 2D target space are comparatively simple to
describe because the only propagating degree of freedom is the tachyon
(and if the latter is switched off the theory becomes topological).
Hence several powerful methods exist to describe the theory
efficiently, e.g.~as matrix models. In particular, strings in
non-trivial backgrounds may be studied in great detail. Here are some
references for further orientation: For the matrix model description
of 2D type 0A/0B string theory
cf.~\cite{Takayanagi:2003sm,Douglas:2003up} (for an extensive review
on Liouville theory and its relation to matrix models and strings in
2D cf.~\cite{Nakayama:2004vk}; some earlier reviews are
refs.~\cite{Ginsparg:1993is}; 
the matrix model for the 2D Euclidean string BH has been constructed in \cite{Kazakov:2000pm}; a study of Liouville theory from the 2D dilaton gravity point of view may be found in \cite{Bergamin:2004pn}).
The low energy effective action for 2D type 0A/0B string theory in the presence of RR fluxes has been studied from various aspects e.g.~in \cite{Klebanov:1998yy,Douglas:2003up,Thompson:2003fz,Gukov:2003yp}.

\subsection{Target space formulation of 2D type 0A/0B string theory}

For sake of definiteness focus will be on 2D type 0A with an equal number $q$ of electric and magnetic D0 branes, but other cases may be studied as well. For vanishing tachyon the corresponding target space action is given by (setting $\ka^2=1$) 
\begin{equation}
\label{eq:0A}
S_{0\rm A}=\frac12 \int \extd^{2}x\, \sqrt{-g}\; \left[e^{-2\phi} \left(  R - 4 \; (\nabla \phi)^{2} + 4b^2\right) - \frac{b^2q^2}{4\pi}\; \right] \, ,
\end{equation}
Obviously, this is a special case of the generic model \eqref{eq:GDT},
with $U,V$ given by the penultimate model in table~\ref{tab:1}, to
which all subsequent considerations --- in particular thermodynamical
issues --- apply.  Note that the dilaton fields $X$ and $\phi$ are
related by $X=\exp{(-2\phi)}$.  The constant $b^2=2/\al^\prime$
defines the physical scale. In the absence of D0 branes, $q=0$, the
model simplifies to the Witten BH, cf.~the third line in table
\ref{tab:1}.

The action defining the tachyon sector up to second order in $\tac$ is
given by (cf.~\eqref{eq:cc2})
\begin{equation}
  \label{eq:sn1}
  S_{\tac} = \frac12 \int\extd^2x\sqrt{-g}\left[ F(X) g^{\mu\nu} (\partial_\mu\tac)(\partial_\nu\tac) + f(X,\tac)\right]\,,
\end{equation}
with
\begin{equation}
  \label{eq:sn22}
  F(X)=X\,,\quad f(\tac,X) = b^2\tac^2\left(X-\frac{q^2}{2\pi}\right)\,.
\end{equation}
The total action is $S_{\rm 0A}+S_{\tac}$. 

\subsection{Exact string Black Hole}

The exact string black hole (ESBH) was discovered by Dijkgraaf,
Verlinde and Verlinde more than a decade ago \cite{Dijkgraaf:1992ba}.
The construction of a target space action for it which does not
display non-localities or higher order derivatives had been an open
problem which could be solved only recently \cite{Grumiller:2005sq}.
There are several advantages of having such an action available: the
main point of the ESBH is its non-perturbative aspect, i.e., it is
believed to be valid to all orders in the string-coupling
$\alpha^\prime$. Thus, a corresponding action captures
non-perturbative features of string theory and allows, among other
things, a thorough discussion of ADM mass, Hawking temperature and
Bekenstein--Hawking entropy of the ESBH which otherwise requires some
ad-hoc assumption.  Therefore, we will devote some space to its
description. 
  At the perturbative level actions approximating
  the ESBH are known: to lowest order in $\alpha^\prime$ one has
  \eqref{eq:0A} with $q=0$.  Pushing perturbative considerations
  further Tseytlin was able to show that up to 3 loops the ESBH is
  consistent with sigma model conformal invariance
  \cite{Tseytlin:1991ht}. In the strong coupling regime the ESBH
  asymptotes to the Jackiw--Teitelboim model \cite{Teitelboim:1983ux}.
  The exact conformal field theory methods used in
  \cite{Dijkgraaf:1992ba}, based upon the
  $\mathrm{SL}(2,\mathbb{R})/\mathrm{U}(1)$ gauged
  Wess--Zumino--Witten model, imply the dependence of the ESBH
  solutions on the level $k$. A different (somewhat more direct)
  derivation leading to the same results for dilaton and metric was
  presented in \cite{Tseytlin:1992ri}.  For a comprehensive history
  and more references \cite{Becker:1994vd} may be consulted.

In the notation of
\cite{Kazakov:2001pj} for Euclidean signature the line element of
the ESBH is given by
\begin{equation}
\extd s^2=f^2(x)\extd\tau^2+\extd x^2\,, \label{eq:dvv1}
\end{equation}
with
\begin{equation}
f(x)=\frac{\tanh(bx)}{\sqrt{1-p\tanh^2(bx)}}\,. \label{eq:dvv2}
\end{equation}
Physical scales are adjusted by the parameter $b\in\mathbb{R}^+$
which has dimension of inverse length. The corresponding
expression for the dilaton,
\begin{equation}
\phi=\phi_0-\ln\cosh(bx)-\frac14\,\ln\left(1-p\tanh^2(bx)\right),
\label{eq:dvv3}
\end{equation}
contains an integration constant $\phi_0$. Additionally, there are
the following relations between constants, string-coupling
$\al^\prime$, level $k$ and dimension $D$ of string target space:
\begin{equation}
\alpha^\prime b^2=\frac{1}{k-2}\,,\qquad
p:=\frac{2}{k}=\frac{2\alpha^\prime b^2}{1+2\alpha^\prime
b^2}\,,\qquad D-26+6\alpha^\prime b^2=0\,.
\label{eq:dvv4}
\end{equation}
For $D=2$ one obtains $p=\frac89$, but like in the original work
\cite{Dijkgraaf:1992ba} we will treat general values of
$p\in(0;1)$ and consider the limits $p\to 0$ and $p\to 1$
separately: for $p=0$ one recovers the Witten BH geometry; for
$p=1$ the Jackiw--Teitelboim model is obtained. Both limits
exhibit singular features: for all $p\in(0;1)$ the solution is
regular globally, asymptotically flat and exactly one Killing
horizon exists. However, for $p=0$ a curvature singularity
(screened by a horizon) appears and for $p=1$ space-time fails to
be asymptotically flat. In the present work exclusively the
Minkowskian version of (\ref{eq:dvv1})
\begin{equation}
\extd s^2=f^2(x)\extd\tau^2-\extd x^2\,, \label{eq:dvv5}
\end{equation}
will be needed. The maximally extended space-time of this geometry has
been studied in \cite{Perry:1993ry}. Winding/momentum mode duality
implies the existence of a dual solution, the Exact String Naked
Singularity (ESNS), which can be acquired most easily by replacing
$bx\to bx+i\pi/2$, entailing in all formulas above the substitutions
$\sinh\to i\cosh$, $\cosh\to i\sinh$.

After it had been realized that the nogo result
of \cite{Grumiller:2002md} may be circumvented without introducing
superfluous physical degrees of freedom by adding an
abelian $BF$-term, a straightforward reverse-engineering procedure
allowed to construct uniquely a target space action of the form
\eqref{eq:GDT}, supplemented by aforementioned $BF$-term,  
\begin{equation}
  \label{eq:ESBHaction}
    S_{\rm ESBH}=-\int \left[X_aT^a+X_{\rm ESBH}R+\epsilon\left(X^+X^- U_{\rm ESBH} + V_{\rm ESBH}\right)\right] - \int BF\,,
\end{equation}
where $B$ is a scalar field and $F=\extd A$ an abelian field strength 2-form.
{\em Per constructionem} $S_{\rm ESBH}$ reproduces as classical solutions
precisely \eqref{eq:dvv2}--\eqref{eq:dvv5} not only locally
but globally. A similar action has been constructed for the ESNS.
The relation $(X-\gamma)^2 = \arcsinh^2{\gamma}$
in conjunction with the definition $\gamma:=\exp{(-2\phi)}/B$
may be used to express the auxiliary dilaton field $X$ entering the action \eqref{eq:GDT} in terms
of the ``true'' dilaton field $\phi$ and the auxiliary field $B$.
The two branches of the square root function correspond to the
ESBH (main branch) and the ESNS (second branch), respectively:
\begin{equation}
  \label{eq:solutionofESBH2.5}
  X_{\rm ESBH} = \gamma + \arcsinh{\gamma}\,,\qquad X_{\rm ESNS} = \gamma - \arcsinh{\gamma}\,.
\end{equation}
The potentials read \cite{Grumiller:2005sq}
\begin{equation}
  \label{eq:solutionofESBH3}
V_{\rm ESBH}=-2b^2\gamma\,,\quad U_{\rm ESBH}= -\frac{1}{\gamma N_{+}(\gamma)}\,, \qquad V_{\rm ESNS}=-2b^2\gamma\,,\quad  U_{\rm ESNS}= -\frac{1}{\gamma N_{-}(\gamma)}\,,
\end{equation}
with 
\begin{equation}
  \label{eq:solutionofESBH4}
  N_\pm(\gamma)=1+\frac{2}{\gamma}\left(\frac{1}{\gamma}\pm\sqrt{1+\frac{1}{\gamma^2}}\right).
\end{equation}
Note that $N_+N_-=1$.
The conformally invariant combination \eqref{eq:wI}, 
\begin{equation}
  \label{eq:wESBHNS}
  w_{\rm ESBH} = -b \left(1+\sqrt{\ga^2+1}\right)\,,\qquad w_{\rm ESNS} = -b \left(1-\sqrt{\ga^2+1}\right)\,,
\end{equation}
of the potentials shows that the ESBH/ESNS is a Minkowskian ground state model, $w\propto 1/I$.
\begin{figure}
\centering \epsfig{file=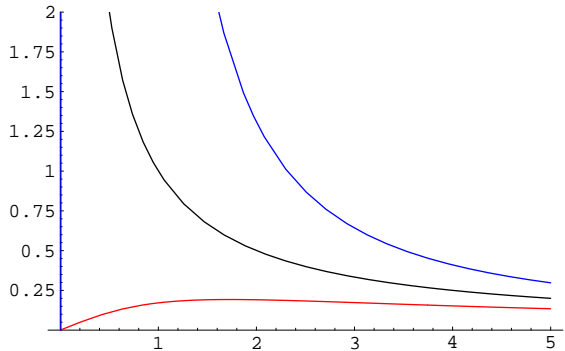,width=.5\linewidth}
\caption{The potentials $U(\ga)$ for the ESNS, the Witten BH and the ESBH.}
\label{fig1}
\end{figure}
In figure \ref{fig1} the potential $U$ is plotted as function of the
auxiliary dilaton $\ga$. The lowest branch is associated with the
ESBH, the one on top with the ESNS and the one in the middle with the
Witten BH (i.e., the third entry in table \ref{tab:1}). The regularity
of the ESBH is evident, as well as the convergence of all three
branches for $\ga\to\infty$, encoding (T-)self-duality of the Witten BH.
For small values of the dilaton the discrepancy between the ESBH, the
ESNS and the Witten BH is very pronounced. Note that $U$ remains
bounded globally only for the ESBH, concurring with the absence of a
curvature singularity.

The two constants of motion --- mass and charge --- may be parameterized
by $k$ and $\phi_0$, respectively. Thus, the level $k$ is not fixed a
priori but rather emerges as a constant of motion, namely essentially
the ADM mass. A rough interpretation of this --- from the stringy point
of view rather unexpected --- result has been provided in
\cite{Grumiller:2005sq} and coincides with a similar one in
\cite{Kazakov:2001pj}. 
There is actually a physical reason why $k$ defines the
  mass: in the presence of matter the conservation equation $\extd M
  =0$ (with $M$ from \eqref{eq:c}) acquires a matter contribution, $\extd M = W^{(m)}$, where
  $W^{(m)}=\extd {\mathcal C}^{(m)}$ is an exact 1-form defined by the energy-momentum tensor
  (cf.~section 5 of \cite{Grumiller:2002nm} or
  \cite{Kummer:1995qv}). In a nutshell, the addition of matter
  deforms the total mass which now consists of a geometric and a
  matter part, $M$ and ${\mathcal C}^{(m)}$,
  respectively. Coming back to the ESBH, the interpretation of $k$ as
  mass according to the preceding discussion implies that the addition
  of matter should ``deform'' $k$. But this is precisely what happens:
  adding matter will in general change the central charge and hence
  the level $k$. Thus, from an intrinsically 2D dilaton gravity point
  of view the interpretation of $k$ as mass is not only possible but
  favored.

It could be interesting to generalize the target space action of 2D type 0A/0B, \eqref{eq:0A}, as to include the non-perturbative corrections implicit in the ESBH by adding \eqref{eq:sn1} (not necessarily with the choice \eqref{eq:sn22}) to the ESBH action \eqref{eq:ESBHaction}. However, it is not quite clear how to incorporate the term from the D0 branes --- perturbatively one should just add $b^2q^2/8\pi$ to $V$ in \eqref{eq:solutionofESBH3}, but non-perturbatively this need not be correct. More results and speculations concerning applications of the ESBH action can be found in \cite{Grumiller:2005sq}.

\section{Black Holes}\label{se:3}

BHs are fascinating objects, both from a theoretical and an
experimental point of view
\cite{Frolov:1998}. 
Many of the features which are generic for BHs are already exhibited
by the simplest members of this species, the Schwarzschild and
Reissner-Nordstr\"om BHs (sometimes the Schwarzschild BH even is
dubbed as ``Hydrogen atom of General Relativity''). Since both of
them, after integrating out the angular part, belong to the class of
2D dilaton gravity models (the first and twelfth model in table
\ref{tab:1}), the study of \eqref{eq:FOG} at the classical,
semi-classical and quantum level is of considerable importance for the
physics of BHs.



\subsection{Classical analysis}

In section \ref{se:sol} it has been recalled briefly how to obtain all
classical solutions in basic EF patches, \eqref{eq:EF}. By looking at
the geodesics of test particles and completeness properties it is
straightforward to construct all Carter-Penrose diagrams for a generic
model \eqref{eq:FOG} (or, equivalently, \eqref{eq:GDT}). For a
detailed description of this algorithm
cf.~\cite{Klosch:1996fi,Klosch:1996qv,Grumiller:2002nm} and references
therein.

\subsection{Thermodynamics}\label{se:td}

\paragraph{Mass} 
The question of how to define ``the'' mass in theories of gravity is
notoriously cumbersome. A nice clarification for ${\rm D}=4$ is
contained in \cite{Faddeev:1982id}. The main conceptual point is that
any mass definition is meaningless without specifying 1.~the ground
state space-time with respect to which mass is being measured and
2.~the physical scale in which mass units are being measured.
Especially the first point is emphasized here. In addition to being
relevant on its own, a proper mass definition is a pivotal ingredient
for any thermodynamical study of BHs. Obviously, any
mass-to-temperature relation is meaningless without defining the
former (and the latter). For a large class of 2D dilaton gravities
these issues have been resolved in \cite{Liebl:1997ti}. One of the key
ingredients is the existence \cite{Frolov:1992xx,Mann:1993yv} of a
conserved quantity \eqref{eq:c} which has a deeper explanation in the
context of first order gravity \cite{Grosse:1992vc} and Poisson-sigma
models \cite{Schaller:1994es}. It establishes the necessary
prerequisite for all mass definitions, but by itself it does not yet
constitute one. Ground state and scale still have to be defined.
Actually, one can take $M$ from \eqref{eq:c} provided the two
ambiguities from integration constants in \eqref{eq:wI} are fixed
appropriately. This is described in detail in appendix A of
\cite{Grumiller:2004wi}. In those cases where this notion makes sense
$M$ then coincides with the ADM mass.

\paragraph{Hawking temperature} There are many ways to calculate the
Hawking temperature, some of them involving the coupling to matter
fields, some of them being purely geometrical. Because of its
simplicity we will restrict ourselves to a calculation of the
geometric Hawking temperature as derived from surface gravity
(cf.~e.g.~\cite{Wald:1999vt}). If defined in this way it turns out to
be independent of the conformal frame.  However, it should be noted
that identifying Hawking temperature with surface gravity is somewhat
naive for space-times which are not asymptotically flat. But the
difference is just a redshift factor and for quantities like entropy
or specific heat actually \eqref{eq:ht} is the relevant quantity as it
coincides with the period of Euclidean time
(cf.~e.g.~\cite{Gibbons:1994cg}).  Surface gravity can be calculated
by taking the normal derivative $\extd/\extd X$ of the Killing norm
(cf.~\eqref{eq:EF}) evaluated on one of the Killing horizons $X=X_h$,
where $X_h$ is a solution of \eqref{eq:horizon}, thus yielding
\begin{equation}
  \label{eq:ht}
  T_H =  \frac{1}{2\pi} \Big|w'(X) \Big|_{X=X_h}\,.  
\end{equation}
The numerical prefactor in \eqref{eq:ht} can be changed e.g.~by a
redefinition of the Boltzmann constant. It has been chosen in
accordance with refs.~\cite{Kummer:1999zy,Grumiller:2002nm}.

\paragraph{Entropy} 
In 2D dilaton gravity there are various ways to calculate the
Beken\-stein-Hawking
entropy. 
Using two different methods (simple thermodynamical considerations,
i.e., $\extd M=T\extd S$, and Wald's Noether charge technique
\cite{Wald:1993nt}) Gegenberg, Kunstatter and Louis-\-Martinez were
able to calculate the entropy for rather generic 2D dilaton gravity
\cite{Gegenberg:1995pv}: entropy equals the dilaton field evaluated at
the Killing horizon,
\begin{equation}
  \label{eq:entropy}
  S = 2\pi X_h\,.
\end{equation}
There exist various ways to count the microstates by appealing to the
Cardy formula \cite{Bloete:1986qm} and to recover the result
\eqref{eq:entropy}. However, the true nature of these microstates
remains unknown in this approach, which is a challenging open problem.
Many different proposals have been made
\cite{Strominger:1996sh}. 

\paragraph{Specific heat}
By virtue of $C_s=T\extd S/\extd T$ the specific heat reads
\begin{equation}
  \label{eq:cs}
  C_s = 2\pi \left. \frac{w'}{w''}\right|_{X=X_h}=\ga_S \,T_H\,,
\end{equation}
with $\ga_S=4\pi^2\,\sign{(w'(X_h))}/w''(X_h)$.  Because it is
determined solely by the conformally invariant combination of the
potentials, $w$ as defined in \eqref{eq:wI}, specific heat is
independent of the conformal frame, too. On a curious sidenote it is
mentioned that \eqref{eq:cs} behaves like an electron gas at low
temperature with Sommerfeld constant $\ga_S$ (which in the present
case may have any
sign). 
If $C_s$ is positive and $C_s T^2\gg 1$ one may calculate logarithmic
corrections to the canonical entropy from thermal fluctuations and
finds \cite{Grumiller:2005vy}
\begin{equation}
  \label{eq:entropy6}
  S_{\rm can} = 2\pi X_h + \frac32\,\ln{\Big|w'(X_h) \Big|} - \frac12\,\ln{\Big|w''(X_h) \Big|} + \dots\,.
\end{equation}  


\paragraph{Hawking-Page like phase transition} In their by now classic
paper on thermodynamics of BHs in $AdS$, Hawking and Page found a
critical temperature signalling a phase transition between a BH phase
and a pure $AdS$ phase \cite{Hawking:1982dh}. This has engendered much
further research, mostly in the framework of the $AdS$/CFT
correspondence (for a review cf.~\cite{Aharony:1999ti}). This
transition is displayed most clearly by a change of the specific heat
from positive to negative sign: for Schwarzschild-$AdS$ (cf.~the
thirteenth entry in table \ref{tab:1}) the critical value of $X_h$ is
given by $X_h^c=\ell^2/3$. For $X_h>X_h^c$ the specific heat is
positive, for $X_h<X_h^c$ it is negative.\footnote{Actually, in the
  original work \cite{Hawking:1982dh} Hawking and Page did not invoke
  the specific heat directly. The consideration of the specific heat
  as an indicator for a phase transition is in accordance with the
  discussion in \cite{Brown:1994gs}.} By analogy, a similar phase
transition may be expected for other models with corresponding
behavior of $C_s$. Interesting speculations on a phase transition at
the Hagedorn temperature $T_h=k/(2\pi)$ induced by a tachyonic
instability have been presented recently in the context of 2D type 0A
strings (cf.~the penultimate model in table \ref{tab:1}) by Olsson
\cite{Olsson:2005en}. From equation (22) of that work one can check
easily that indeed the specific heat (at fixed $q$),
$C_s=(q^2/8)(T/T_h)/(1-T/T_h)$, changes sign at $T=T_h$.


\subsection{Semi-classical analysis}

After the influential CGHS paper \cite{Callan:1992rs} there has been a
lot of semi-classical activity in 2D, most of which is summarized in
\cite{Harvey:1992xk,Kummer:1999zy,Grumiller:2002nm}. In many
applications one considers \eqref{eq:GDT} coupled to a scalar field
\eqref{eq:cc2} with $F=\rm const.$ (minimal coupling). Technically,
the crucial ingredient for 1-loop effects is the Weyl anomaly
(cf.~e.g.~\cite{Duff:1993wm}) $<T^\mu_\mu> = R/(24\pi)$, which ---
together with the semi-classical conservation equation
$\nabla_\mu<T^{\mu\nu}>=0$ --- allows to derive the flux component of
the energy momentum tensor after fixing some relevant integration
constant related to the choice of vacuum (e.g.~Unruh, Hartle-Hawking
or Boulware). This method goes back to Christensen and Fulling
\cite{Christensen:1977jc}. For non-minimal coupling, e.g.~$F\propto
X$, there are some important modifications --- for instance, the
conservation equation no longer is valid but acquires a right hand
side proportional to $F'(X)$. The first calculation of the conformal
anomaly in that case has been performed by Mukhanov, Wipf and Zelnikov
\cite{Mukhanov:1994ax}. It has been confirmed and extended e.g.~in
\cite{Chiba:1997ex}.

\subsection{Long time behavior}

The semi-classical analysis, while leading to interesting results, has
the disadvantage of becoming unreliable as the mass of the evaporating
BH drops to zero. The long time behavior of an evaporating BH presents
a challenge to theoretical physics and touches relevant conceptual
issues of quantum gravity, such as the information paradox. There are
basically two strategies: top-down, i.e., to construct first a full
quantum theory of gravity and to discuss BH evaporation as a
particular application thereof, and bottom-up, i.e., to sidestep the
difficulties inherent to the former approach by invoking
``reasonable'' ad-hoc assumptions. The latter route has been pursued
in \cite{Grumiller:2003hq}. A crucial technical ingredient has been
Izawa's result \cite{Izawa:1999ib} on consistent deformations of 2D BF
theory, while the most relevant physical assumption has been
boundedness of the asymptotic matter flux during the whole evaporation
process. Together with technical assumptions which can be relaxed, the
dynamics of the evaporating BH has been described by means of
consistent deformations of the underlying gauge symmetries with only
one important deformation parameter. In this manner an attractor
solution, the endpoint of the evaporation process, has been found
(cf.~the eighth model in table \ref{tab:1}).

Ideologically, this resembles the exact renormalization group
approach, cf.~e.g. \cite{Lauscher:2001ya,Bonanno:2006eu} and
references therein, which is based upon Weinberg's idea of
``asymptotic safety''.%
\footnote{In the present context also \cite{Niedermaier:2003fz} should
  be mentioned.}  There are, however, several conceptual and technical
differences, especially regarding the truncation of ``theory space'':
in 4D a truncation to EH plus cosmological constant, undoubtedly a
very convenient simplification, may appear to be somewhat ad-hoc,
whereas in 2D a truncation to \eqref{eq:FOG} comprises not only
infinitely many different theories, but essentially\footnote{Actually,
  one should replace in \eqref{eq:FOG} the term $X^+X^-U(X)+V(X)$ by
  $\mathcal{V}(X^+X^-,X)$. However, only \eqref{eq:FOG} allows for
  standard supergravity extensions \cite{Bergamin:2002ju}.} {\em all}
theories with the same field content as \eqref{eq:FOG} and the same
kind of local symmetries (Lorentz transformations and
diffeomorphisms).

The global structure of an evaporating BH can also be studied, and
despite of the differences between various approaches there seems to
be partial agreement on it,
cf.~e.g.~\cite{Frolov:1981mz,Parikh:1998ux,Grumiller:2003hq,Ashtekar:2005cj,Hayward:2005gi,Bonanno:2006eu}.
The crucial insight might be that a BH in the mathematical sense
(i.e., an event horizon) actually never forms, but only some trapped
region, cf.~figure 5 in \cite{Hayward:2005gi}.

\subsection{Killing horizons kill horizon degrees}\label{se:kill}

As pointed out by Carlip \cite{Carlip:2004mn}, the fact that very
different approaches to explain the entropy of BHs nevertheless agree
on the result urgently asks for some deeper explanation. 
Carlip's suggestion was to consider an
underlying symmetry, somehow attached to the BH horizon, as the key
ingredient, and he noted that requiring the presence of a horizon
imposes constraints on the physical phase space. Actually, the change
of the phase-space structure due to a constraint which imposes the
existence of a horizon in space-time is an issue which is of
considerable interest by itself.

In a recent work \cite{Bergamin:2005pg} we could show that the
classical physical phase space is smaller as compared to the generic
case if horizon constraints are imposed. Conversely, the number of
gauge symmetries is larger for the horizon scenario. In agreement with
a conjecture by 't Hooft \cite{'tHooft:2004ek}, we found that physical
degrees of freedom are converted into gauge degrees of freedom at a
horizon. We will now sketch the derivation of this result briefly for
the action \eqref{eq:FOG} which differs from the one used in
\cite{Bergamin:2005pg} by a (Gibbons-Hawking) boundary term. For sake
of concreteness we will suppose the boundary is located at $x^1=\rm
const$. Consistency of the variational principle then requires
\begin{equation}
  \label{eq:newboundary}
  X^+ \de e_0^-  + X^-\de e_0^+  + X \de\om_0 = 0
\end{equation}
at the boundary. Note that one has to fix the parallel component of
the spin-connection at the boundary rather than the dilaton field,
which is the main difference to \cite{Bergamin:2005pg}.  The generic
case imposes $\de e_0^\pm=0=\de\om_0$, while a horizon allows the
alternative prescription $\de e_0^-=X^-=0=\de\om_0$. One can now
proceed in the same way as in \cite{Bergamin:2005pg}, i.e., derive the
constraints (the only boundary terms in the secondary constraints are
now $X$ and $X^\pm$, while the primary ones have none) and calculate
the constraint algebra. All primary constraints and the Lorentz
constraint turn out to be first class, even at the boundary, whereas
the Poisson bracket between the two diffeomorphism constraints ($G_2$,
$G_3$ in the notation of \cite{Bergamin:2005pg}) acquires a boundary
term of the form
\begin{equation}
  \label{eq:boundary}
  X(U'X^+X^-+V') + U(X)X^+X^- - V(X)\,.
\end{equation}
Notably, it vanishes only for $V\propto X$ and $U\propto 1/X$,
e.g.~for the second, third and sixth model in table \ref{tab:1}, i.e.,
$(A)dS_2$ ground state
models. 
The boundary constraints for the generic case convert all primary
constraints into second class constraints. The construction of the
reduced phase space works in the same way as in section 6 of
\cite{Bergamin:2005pg}, thus establishing again one physical degree of
freedom ``living on the boundary''. Actually, this had been known
already before \cite{Kuchar:1994zk}.  The horizon constraints,
however, lead to more residual gauge symmetries and to a stronger
fixing of free functions --- in fact, no free function remains and the
reduced phase space is empty. Thus, the physical degree of freedom
living on a generic boundary is killed by a Killing horizon.

It would be interesting to generalize this physics-to-gauge conversion
at a horizon to the case with matter. Obviously, it will no longer be
a Killing horizon, but one can still employ the (trapping) horizon
condition $X^-=0$.

\subsection{Critical collapse}

Critical phenomena in gravitational collapse have been
discovered in the pioneering numerical investigations of
Choptuik \cite{Choptuik:1993jv}. 
He studied a free massless scalar field coupled
to spherically symmetric EH gravity in 4D (the EMKG) 
with sophisticated numerical techniques that allowed him to
analyze the transition in the space of initial data
between dispersion to infinity and the formation of a BH.
Thereby the famous scaling law
\begin{equation}
  \label{eq:choptuik}
  M_{BH} \propto (p-p_\ast)^\gamma\,,
\end{equation}
has been established, where $p\in[0,1]$ is a free parameter
characterizing a one-parameter family of initial data with the
property that for $p<p_\ast$ a BH never forms while for $p>p_\ast$ a
BH always forms with mass $M_{BH}$ determined by \eqref{eq:choptuik}
for $p$ sufficiently close to $p_\ast$. The critical parameter
$p_\ast\in(0,1)$ may be found by elaborate numerical analysis and
depends on the specific family under consideration; but the critical
exponent $\gamma\approx 0.37$ is universal, albeit model dependent.
Other systems may display a different critical behavior, cf.~the
review \cite{Gundlach:1998wm}. The critical solution $p=p_\ast$,
called the ``Choptuon'', in general exhibits remarkable features,
e.g.~discrete or continuous self-similarity and a naked singularity.

Since the original system studied by Choptuik, \eqref{eq:cc22}, is a
special case of \eqref{eq:GDT} (with $U,V$ as given by the first line
in table \ref{tab:1}) coupled to \eqref{eq:cc2}, it is natural to
inquire about generalizations of critical phenomena to arbitrary 2D
dilaton gravity with scalar matter. Indeed, in
\cite{Strominger:1993tt} a critical exponent $\ga=1/2$ has been
derived analytically for the RST model \cite{Russo:1993yh}, a
semi-classical generalization of the CGHS model (cf.~the third line in
table \ref{tab:1}). Later, in \cite{Peleg:1996ce} critical collapse
within the CGHS model has been considered and $\ga\approx 1/2$ has
been found numerically.  
More recently the generalization of the original Choptuik
system to D dimensions has been considered
\cite{Birukou:2002ge,Sorkin:2005vz,Bland:2005kk}. For $3.5\leq D\leq
14$ the approximation $\ga(D)\approx 0.47(1-\exp{(-0.41 D)})$ shows
that $\ga$ increases monotonically\footnote{In \cite{Sorkin:2005vz} a
  maximum in $\ga$ near D=11 has been found.  The most recent study
  suggests it is an artifact of numerics \cite{Bland:2005kk}. Another
  open question concerns the limit ${\rm D}\to 3$: does $\ga$ remain
  finite?} with D. Since formally the CGHS corresponds to the limit
${\rm D}\to\infty$ one may expect that $\ga(D)$ asymptotes to the
value $\ga\approx 1/2$.

In the remainder of this subsection we will establish evolution
equations for generic 2D dilaton gravity with scalar matter, to be
implemented numerically analogously to
\cite{Husa:2000kr,Purrer:2004nq}. In these works for various reasons
Sachs-Bondi gauge has been used. Thus we employ
\begin{equation}
  \label{eq:cc5}
  e_0^+=0\,,\quad e_0^-=1\,,\quad x^0=X\,,
\end{equation}
while the remaining Zweibein components are parameterized as
\begin{equation}
  \label{eq:cc6}
  e_1^-=\al(u,X)\,,\quad e_1^+=I(X)e^{2\beta(u,X)}\,.
\end{equation}
In the gauge \eqref{eq:cc5} with the parameterization \eqref{eq:cc6}
the line element reads
\begin{equation}
  \label{eq:cc7}
  \extd s^2=2I(X)e^{2\be(u,X)}\extd u \left(\extd X+\al(u,X)\extd u\right)\,.
\end{equation}
A trapping horizon emerges either if $\al=0$ or $\be\to\infty$. The
equations of motion may be reduced to the following set:
\begin{align}
& {\rm Slicing\,\,condition:} & \partial_X\al(u,X)&=-e^{2\be(u,X)}w^\prime(X)   \label{eq:eom1}\\
& {\rm Hamiltonian\,\,constraint:} & \partial_X\be(u,X)&=-F(X)(\partial_X\tac(u,X))^2  \label{eq:eom2} \\
& {\rm Klein-Gordon\,\,equation:} & \square\;\tac(u,X)&=0  \label{eq:eom3}
\end{align}
with 
\begin{equation}
  \label{eq:box}
  \square=2\partial_X\partial_u-2\partial_X(\al(u,X)\partial_X)-\frac{F^\prime(X)}{F(X)}\left(2\al(u,X)\partial_X-\partial_u\right)\,.
\end{equation}
These equations should be compared with (2.12a), (2.12b) in
\cite{Husa:2000kr} or with (2.4) (and for the Klein-Gordon equation
also (2.3)) in \cite{Purrer:2004nq}, where they have been derived for
spherically symmetric EH gravity in 4D.  In the present case they are
valid for generic 2D dilaton gravity coupled non-minimally to a free
massless scalar field. Thus, the set of equations
\eqref{eq:eom1}-\eqref{eq:box} is a suitable starting point for
numerical simulations in generic 2D dilaton gravity.  The Misner-Sharp
mass function
\begin{equation}
  \label{eq:cc17}
  m(u,X)=-X^+X^-I(X)-w(X)=-\al(u,X) e^{-2\be(u,X)}-w(X)
\end{equation}
allows to rewrite the condition for a trapped surface as $\al
e^{-2\be}=0$ (cf.~\eqref{eq:horizon} with \eqref{eq:c}).  Thus, as
noted before, either $\al$ has to vanish or $\be\to\infty$; it is the
latter type of horizon that is of relevance for numerical simulations
of critical collapse. One may use the Misner-Sharp function instead of
$\al$ and thus obtains instead of \eqref{eq:eom1}
\begin{equation}
  \label{eq:cc19}
  \partial_X m(u,X)=(m(u,X)+w(X))2F(X)(\partial_X\tac(u,X))^2\,.
\end{equation}
To monitor the emergence of a trapped surface numerically one has to
check whether
\begin{equation}
  \label{eq:trapped}
  m(u_0,X_h)+w(X_h)\approx 0
\end{equation}
is fulfilled to a certain accuracy at a given retarded time $u_0$; the
quantity $X_h$ corresponds to the value of the dilaton field at the
horizon. By analogy to (2.16) of \cite{Purrer:2004nq} one may now
introduce a compactified ``radial'' coordinate, e.g.~$X/(1+X)$,
although there may be more convenient choices.

As a consistency check the original Choptuik system in the current
notation will be reproduced. We recall that \eqref{eq:cc22} describes
the EMKG.
Using $\extd r=I(X)\extd X$ the evolution equations for geometry read:
\begin{align} 
  \label{eq:cc24}
  & \partial_r\beta=\frac{\ka}{2}r(\partial_r\tac)^2 \\
  \label{eq:cc25}
  & \partial_r\alpha=\la^2e^{2\be}
\end{align}
They look almost the same as (2.4) in \cite{Purrer:2004nq}. The
coupling constant $\ka$ just has to be fixed appropriately in
\eqref{eq:cc24} (i.e.~$\ka=4\pi$). Also, the scaling constant $\la$
must be fixed. Note that the line element reads
\begin{equation}
  \label{eq:cc26}
  \extd s^2=2\frac{2}{r}e^{2\be(u,X(r))}\extd u \left(\extd r \frac{r}{2}+\al(u,X(r))\extd u\right)= 2e^{2\be}\extd u \left(\extd r + \frac{2\al}{r}\extd u\right)
\end{equation}
This shows that $\be$ here really coincides with $\be$ in
\cite{Purrer:2004nq} and $\al$ here coincides, up to a numerical
factor, with $V$ there (and there are some signs due to different
conventions). 



\subsection{Quasinormal modes}

The term ``quasinormal modes'' refers to some set of modes with a
complex frequency, associated with small perturbations of a BH.  For
$U=0$ and monomial $V$ in \cite{Kettner:2004aw} quasinormal modes
arising from a scalar field, \eqref{eq:cc2} with $f=0$ and $F\propto
X^p$, have been studied in the limit of high damping by virtue of the
``monodromy approach'', and the relation
\begin{equation}
  \label{eq:qnm}
  e^{\om /T_H}=-(1+2\cos{(\pi(1-p))})
\end{equation}
for the frequency $\om$ has been found ($T_H$ is Hawking temperature
as defined in \eqref{eq:ht}). Minimally coupled scalar fields ($p=0$)
lead to the trivial result $\om/T_H=2\pi i n$. High damping implies
that the integer $n$ has to be large. For the important case of $p=1$
(relevant for the first and fifth entry in table \ref{tab:1}) one
obtains from \eqref{eq:qnm}
\begin{equation}
  \label{eq:qnmss}
  \frac{\om}{T_H}=2\pi i \left(n+\frac12\right) + \ln{3}\,.
\end{equation}
The result \eqref{eq:qnmss} coincides with the one obtained for the
Schwarzschild BH with 4D methods, both numerically \cite{Nollert:1993}
and analytically \cite{Motl:2002hd}. Moreover, consistency with ${\rm
  D}>4$ is found as well \cite{Motl:2003cd}. This shows that the 2D
description of BHs is reliable also with respect to highly damped
quasinormal modes.

\subsection{Solid state analogues} 

BH analogues in condensed matter systems go back to the seminal paper
by Unruh \cite{Unruh:1980cg}. Due to the amazing progress in
experimental condensed matter physics, in particular Bose-Einstein
condensates, in the past decade the subject of BH analogues has
flourished, cf.~e.g.~\cite{Novello:2002qg} and references therein.

In some cases the problem effectively reduces to 2D. It is thus
perhaps not surprising that an analogue system for the
Jackiw-Teitelboim model has been found \cite{Fedichev:2003id} for a
cigar shaped Bose-Einstein condensate. More recently this has led to
some analogue 2D activity \cite{Balbinot:2004da}. 
Note,
  however, that some issues, like the one of backreaction, might not
  be modelled very well by an effective action method
  \cite{Schutzhold:2005ex}.
Indeed, 2D dilaton gravity with matter
could be of interest in this context, because these systems might
allow not just kinematical but dynamical equivalence, i.e., not only
the fluctuations (e.g.~phonons) behave as the corresponding
gravitational ones (e.g.~Hawking quanta), but also the background
dynamics does (e.g.~the flow of the fluid or the metric,
respectively). Such a system would be a necessary pre-requisite to
study issues of mass and entropy in an analogue context. At least for
static solutions this is possible \cite{Grumiller:Dresden}, but of
course the non-static case would be much more interesting. Alas, it is
not only more interesting but also considerably more difficult, and a
priori there is no reason why one should succeed in finding a fully
fledged analogue model of 2D dilaton gravity with matter. Still, one
can hope and try.

\section{Geometry from matter}\label{se:4}

In first order gravity \eqref{eq:FOG} coupled to scalar \eqref{eq:cc2}
or fermionic \eqref{eq:cc2ferm} matter the geometry can be quantized
exactly: after analyzing the constraints, fixing EF gauge
\eq{(\om_0,e_0^-,e_0^+) = (0,1,0)}{eq:EFgauge}
and constructing a BRST invariant Hamiltonian, 
the path integral can be evaluated exactly and a
(nonlocal) effective action is obtained 
\cite{Kummer:1998zs}. Subsequently the matter fields can be quantized
by means of ordinary perturbation theory. To each order all
backreactions are included automatically by this procedure.

\begin{wrapfigure}{l}{3cm}
\epsfig{file=virtualBH.epsi,height=6cm}
\caption{VBH}
\label{fig:cp}
\end{wrapfigure}
Although geometry has been integrated out exactly, it can be recovered
off-shell in the form of interaction vertices of the matter fields,
some of which resemble virtual black holes (VBHs)
\cite{Grumiller:2000ah,Grumiller:2001rg,Grumiller:2002dm}.
This metamorphosis of geometry however does not take place in the
matterless case \cite{Kummer:1997hy}, where the quantum
effective action coincides with the classical action in EF gauge. We
hasten to add that one should not take this off-shell geometry at face
value --- this would be like over-interpreting the role of virtual
particles in a loop diagram. But the simplicity of such geometries and
the fact that all possible configurations are summed over are both
nice qualitative features of this picture.

A Carter-Penrose diagram of a typical VBH configuration is depicted in
figure \ref{fig:cp}.  The curvature scalar of such effective geometries 
is discontinuous and even has a $\de$-peak. A typical effective line
element (for the EMKG) reads
\eq{(\extd s)^2 = 2\extd r \extd u + \left(1 -
    \theta(r_y-r)\de(u-u_y)\left(\frac{2 m}{r} + a r - d\right)\right) (\extd
  u)^2\,,}{eq:VBHlinelement}
It obviously has a Schwarzschild part with $r_y$-dependent ``mass''
$m$ and a Rindler part with $r_y$-dependent ``acceleration'' $a$, both
localized on a lightlike cut. This geometry is nonlocal in the sense
that it depends not just on the coordinates $r,u$ but additionally on
a second point
$r_y,u_y$. 
While the off-shell geometry \eqref{eq:VBHlinelement} is highly gauge
dependent, the ensuing S-matrix --- the only {\em physical} observable
in this context \cite{Kummer:2005tx} --- appears to be
gauge independent, although a formal proof of this statement,
e.g.~analogously to \cite{Kummer:2001ip}, is lacking.

\subsection{Scalar matter}

After integrating out geometry and the ghost sector (for $f(X,\tac)=
0$), the effective Lagrangian ($w$ is defined in \eqref{eq:wI})
\eq{\cL^{\rm eff}_\tac = F(\hat{X})(\partial_0 \tac)(\partial_1 \tac) - w'(\hat{X})+\rm sources
}{eq:Leffscalar}
contains the quantum version of the dilaton field
$\hat{X}=\hat{X}(\nabla_0^{-2}(\partial_0\tac)^2)$, depending
non-locally on $\tac$.  The quantity $\hat{X}$ solves the equation of
motion of the classical dilaton field, with matter terms and external
sources for the geometric variables in EF gauge. The simplicity of
\eqref{eq:Leffscalar} is in part due to the gauge choice
\eqref{eq:EFgauge} and in part due to the linearity of the gauge fixed
Lagrangian in the remaining gauge field components, thus producing
delta-functionals upon path integration.
 
\begin{figure}
\centering \epsfig{file=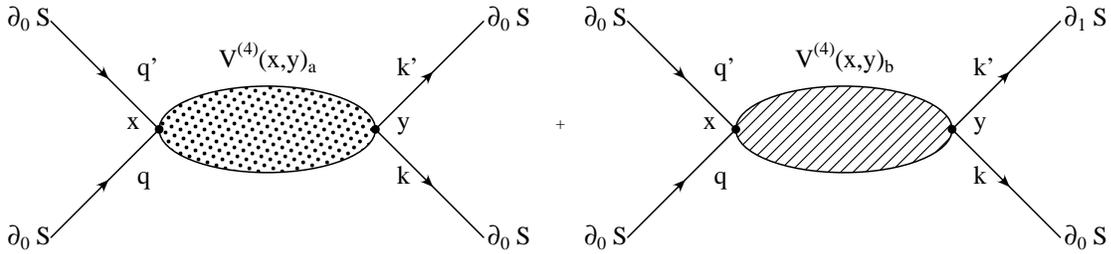,width=\linewidth}
\caption{Non-local 4-point vertices} \label{fig:feyn}
\end{figure}

In principle, the interaction vertices can be extracted by expanding
the nonlocal effective action in a power series of the scalar field
$\tac$. However, this becomes cumbersome already at the $\tac^4$
level. Fortunately, the localization technique introduced in
\cite{Kummer:1998zs} simplifies the calculations considerably. It
relies on two observations: First, instead of dealing with complicated
nonlocal kernels one may solve corresponding differential equations after
imposing asymptotic
conditions on the solutions. Second, instead of taking the $n$-th
functional derivative of the action with respect to bilinear
combinations of $\tac$, the matter fields may be localized at $n$
different space-time points, which mimics the effect of functional
differentiation. For tree-level calculations it is then sufficient to
solve the classical equations of motion in the presence of these
sources, which is achieved most easily via appropriate
matching conditions.
It turns out (as anticipated from \eqref{eq:VBHlinelement}) that the
conserved quantity \eqref{eq:c} is discontinuous for a VBH.  This
phenomenon is generic
\cite{Grumiller:2002dm}.\footnote{With the exception of scattering
  trivial models, cf.~the eleventh entry in table \ref{tab:1}.} 
The corresponding Feynman diagrams are contained in figure \ref{fig:feyn}.\footnote{The scalar field $\tac$ is denoted by $S$ in these graphs.} 
For free, massless, non-minimally coupled scalars ($F\neq \rm const.$)
both the symmetric and the non-symmetric 4-point vertex
\begin{equation}
  \label{eq:added2}
  V^{(4)}=\int \extd^2 x
\extd^2 y (\partial_0 \tac)^2_x \left[V_{\rm a}(x,y) (\partial_0
  \tac)^2_y + V_{\rm b}(x,y) (\partial_0 \tac)_y (\partial_1 \tac)_y
\right]
\end{equation}
are given in \cite{Grumiller:2002dm}, and have the following
properties:
\begin{enumerate}
\setlength{\itemsep}{-1pt}
\item They are local in one coordinate (e.g.~containing $\de(x^1-y^1)$) and nonlocal in the other.
\item They vanish in the local limit ($x^0\to y^0$). Additionally, $V_{\rm b}$ vanishes for minimal coupling $F={\rm const.}$
\item The symmetric vertex depends only on the conformal invariant
  combination $w(X)$ and the asymptotic value $M_\infty$ of
  \eqref{eq:c}. The non-symmetric one is independent of $U$, $V$ and
  $M_\infty$. Thus if $M_\infty$ is fixed in all conformal frames,
  both vertices are conformally invariant.
\item They respect the ${\Bbb Z}_2$ symmetry $F(X) \mapsto - F(X)$.
\end{enumerate}
It should be noted that the class of models with $UV+V'=0$ and $F={\rm
  const}$ (containing the CGHS model, the seventh and eleventh entry in table
\ref{tab:1}) shows ``scattering triviality'', i.e., the classical
vertices vanish, and scattering can only arise from higher order
quantum backreactions. For these models the VBH has no
classically observable consequences, but at 1-loop level physical
observables like the specific heat are modified appreciably
\cite{Grumiller:2003mc}.

The 2D Klein-Gordon equation relevant for the construction of asymptotic states is also conformally invariant.
For minimal coupling it simplifies considerably, and a complete set of
asymptotic states can be obtained explicitly. Since both, asymptotic states and
vertices, only depend on $w(X)$ and $M_\infty$, at tree level
conformal invariance holds nonperturbatively (to all orders in
$\tac$), but it is broken at 1-loop level due to the conformal
anomaly. Because asymptotically geometry does not fluctuate, a
standard Fock space may be built with creation/annihilation operators
$a^\pm(k)$ obeying the standard commutation relations. The S-matrix
for two ingoing ($q,q^\prime$) into two outgoing ($k,k^\prime$)
asymptotic modes is determined by (cf.~\eqref{eq:added2})
\begin{equation}
  \label{eq:added1}
  T(q,q^\prime;k,k^\prime)=\frac12 \langle 0 \left| a^-(k)a^-(k^\prime)V^{(4)}a^+(q)a^+(q^\prime)\right| 0 \rangle\,.
\end{equation}
The simple choice $M_\infty=0$ yields a ``standard QFT vacuum'' $|0\rangle$,
provided the model under consideration has a Minkowskian ground state
(e.g.~the first, third, fifth and last model in table \ref{tab:1}).


For the physically interesting case of the EMKG model such an S-matrix was
obtained in \cite{Fischer:2001vz,Grumiller:2001rg}.
Both the symmetric and the non-symmetric vertex contribute, each
giving a divergent contribution to the S-matrix, but the sum of both
turned out to be finite! The whole calculation is highly nontrivial,
involving cancellations of polylogarithmic terms,
%
%
but at the end giving the surprisingly simple result
\eq{
T(q, q'; k, k') = -\frac{i\ka\de\left(k+k'-q-q'\right)}{2(4\pi)^4
|kk'qq'|^{3/2}} E^3 \tilde{T}\,,
}{eq:scatamp}
with ingoing ($q,q'$) and outgoing ($k,k'$) spatial momenta, total
energy $E=q+q'$,
\eq{
\tilde{T} :=  \frac{1}{E^3}{\Bigg [}\Pi \ln{\frac{\Pi^2}{E^6}} 
+ \frac{1} {\Pi} \sum_{p \in \left\{k,k',q,q'\right\}}p^2 \ln{\frac{p^2}{E^2}} 
\cdot {\Bigg (}3 kk'qq'-\frac{1}{2}
\sum_{r\neq p} \sum_{s \neq r,p}\left(r^2s^2\right){\Bigg )} {\Bigg ]},
}{eq:feynmanamplitude}
and the momentum transfer function $\Pi = (k+k')(k-q)(k'-q)$. The
factor $\tilde{T}$ is invariant under rescaling of the momenta $p
\mapsto ap$, and the whole amplitude transforms monomial like $T
\mapsto a^{-4}T$. It should be noted that due to the non-locality of
the vertices there is just one $\de$-function of momentum conservation
(but no separate energy conservation) present in \eqref{eq:scatamp}.
This is advantageous because it eliminates the problem of ``squared
$\de$-functions'' that is otherwise present in 2D theories of massless
scalar fields (cf.~e.g.~\cite{Balasin:1992nm}). In this sense gravity
acts as a regulator of the theory.

The corresponding differential cross section also reveals interesting
features \cite{Grumiller:2001rg}:
\begin{enumerate}
\item For vanishing $\Pi$ forward scattering poles are present.
\item There is an approximate self-similarity close to the forward
  scattering peaks. Far away from them it is broken, however.
\item It is CPT invariant.
\item An ingoing s-wave can decay into three outgoing ones. Although this may
be expected on general grounds, within the present formalism it is possible
to provide explicit results for the decay rate.
\end{enumerate}
Although it seems straightforward to generalize \eqref{eq:added1} to
arbitrary $n$-point vertices, no such calculation has been attempted
so far. This is related to the fact that the derivation of
\eqref{eq:scatamp} has been somewhat tedious and lengthy. Thus, it
could be worthwhile to find a more efficient way to obtain this
interesting S-matrix element.

\subsection{Fermionic matter}

Recently we considered 2D dilaton gravity \eqref{eq:FOG} coupled to
fermions \eqref{eq:cc2ferm} along the lines of the previous
subsection. The results will be published elsewhere, but we give a
short summary with emphasis on differences to the scalar case.

The constraint analysis for the general case \eqref{eq:cc2ferm} has
been worked out first in \cite{Meyer:2005fz}.  Three first class
constraints generating the two diffeomorphisms and the local Lorentz
symmetry and four well-known second class constraints relating the
four real components of the Dirac spinor to their canonical momenta
are present in the system. As anticipated the Hamiltonian is fully
constrained. After introducing the Dirac bracket the constructions of
the BRST charge and the gauge fixed Hamiltonian are straightforward.
Path integration over geometry is even simpler than in the scalar
case, because the second class constraints are implemented in the path
integral through delta functionals, allowing to integrate out the
fermion momenta. The effective Lagrangian
\beqa\nonumber
\cL^{\rm eff}_{\chi} & = & \frac{i}{\sqrt{2}}F(\hat{X}) (\chid_1
  \plrd{1}\chi_1) \\ \label{eq:Leffferm}
      & & + I(\hat{X})\left(
  \frac{i}{\sqrt{2}} F(\hat{X})(\chid_0 \plrd{0} \chi_0) +
  H(\hat{X})g(\chib\chi) - V(\hat{X})\right) + \rm sources
\eeqa
again depends on the quantum version
$\hat{X}=\hat{X}(\nabla_0^{-2}(\chid_1 \plrd{0}\chi_1))$ of the
dilaton field and exhibits non-locality in the matter field.

Some properties remain the same as compared to previous studies with
scalar matter. For instance, the VBH phenomenon is still present, now
even for the eleventh model in table \ref{tab:1}. In fact, the
conserved quantity \eqref{eq:c} now becomes continuous only for the
trivial case $F(X)=0$. But there are also some notable differences.
For example, the non-selfinteracting system already has three 4-point
vertices, two of them being the symmetric and asymmetric vertices of
the scalar case and a new third one, arising from the first term in
the second line of \eqref{eq:Leffferm}. All vertices show the first
two properties listed above, and the symmetric and non-symmetric ones
also the third one.

The new vertex however does not vanish for minimal coupling, and thus
in contrast to the scalar case there are two vertices present even for
this simple case.  It is not conformally invariant, but rather
transforms additively because it contains a term proportional to
$U(X)$. However, since also the external legs have a conformal weight,
conformal invariance of the tree-level S-matrix still is expected to
hold, despite of the non-invariance of some of the vertices and some
of the asymptotic modes.

At 1-loop level and for minimal coupling conformal symmetry is broken
and, exactly as in the case of scalar matter, the conformal anomaly
can be integrated to the non-local Polyakov action
\cite{Polyakov:1981rd}. This has been applied
e.g.~in\cite{Nojiri:1992st}.  A possible Thirring term can be
reformulated using \eqref{eq:rewritethirring} and integrated by use of
the chiral anomaly, giving a Wess-Zumino \cite{Wess:1971yu}
contribution to the effective action. In this case, a path integral
over the auxiliary vector potential remains, with a highly non-local
self-interaction. Whether this treatment is favourable over treating
the Thirring term directly as an interaction vertex has to be decided
by application.

Another peculiar feature of 2D field theories is bosonization, 
e.g.~the quantum equivalence of the Thirring model and the Sine-Gordon
model, both in flat 1+1 dimensions \cite{Coleman:1974bu}. 
This issue has been addressed recently on
a curved background by Frolov, Kristj{\'a}nsson and Thorlacius
\cite{Frolov:2005ps} to investigate the effect of pair-production on
BH space times in regions of small curvature (as compared to the
microscopic length scale of quantum theory). In the framework of first
order gravity it may be possible to investigate the question of
bosonization even outside this simple framework, since one is able to
integrate out geometry non-perturbatively.

\section{Mathematical issues}\label{se:5}

In the absence of matter many of the interesting features discussed in
the previous three sections are absent: there is no tachyon dynamics,
no Hawking radiation, no interesting semi-classical behavior, no
critical collapse, no quasinormal modes, no relevant solid state
analogue, no scattering processes and no reconstruction of geometry
from matter. Nevertheless, some basic features remain, like the global
structure of the classical solutions or the physics-to-gauge
conversion mentioned in section \ref{se:kill}. Mathematically,
however, the absence of matter bears some attractiveness and reveals
beautiful structures responsible for the classical integrability of
\eqref{eq:FOG}. They may allow some relevant generalizations of
\eqref{eq:FOG}, e.g.~in the context of non-commutative gravity.
 
\subsection{Remarks on the Einstein-Hilbert action in 2D}\label{se:EH2D}

In 2D the Einstein tensor vanishes identically for any 2D metric and
thus conveys no useful information. Similarly, the 2D EH action,
supplemented appropriately by boundary and corner terms, just counts
the number of holes of a compact Riemannian manifold,
cf.~e.g.~\cite{Guillemin:1974}. Thus, as compared to \eqref{eq:GDT} or
\eqref{eq:FOG} the study of ``pure'' 2D gravity, i.e., without
coupling to a dilaton field, is of rather limited interest.
If one adds a cosmological constant term one may study
  quantum gravity in 2D by means of dynamical triangulations,
  cf.~e.g.~\cite{Ambjorn:1998fd} and references therein. The EH part
  of the action plays no essential role, however.

It is possible to consider EH gravity in $2+\eps$ dimensions, an idea
which seems to go back to \cite{Gastmans:1977ad}. After taking the
limit $\eps\to 0$ in a specific way \cite{Mann:1992ar} one obtains
again a dilaton gravity model \eqref{eq:GDT} with $V=0$ and $U=\rm
const.$ (cf.~the eleventh model in table \ref{tab:1}). That such a
limit can be very subtle has been shown recently by Jackiw
\cite{Jackiw:2005su} in the context of Weyl invariant scalar field
dynamics: if one simply drops the EH term in equation (3.5) of that
work the Liouville model is obtained (cf.~the tenth model in table
\ref{tab:1}), but Weyl invariance is lost.

\subsection{Relations to 3D: Chern-Simons and BTZ}\label{se:BTZ}
  
The gravitational Chern-Simons term
\cite{Deser:1982wh} and the 3D BTZ BH
\cite{Banados:1992wn} have inspired a lot of further
research. Here we will focus on relations to \eqref{eq:GDT} and
\eqref{eq:FOG}: dimensional reduction of the BTZ to 2D has been
performed in \cite{Achucarro:1993fd}, cf.~the fifteenth model in table
\ref{tab:1}. A reduction of the gravitational Chern-Simons term from
3D to 2D has been performed in \cite{Guralnik:2003we}, cf.~the
sixteenth model in table \ref{tab:1}. Recently \cite{Sahoo:2006vz},
such reductions have been exploited to calculate the entropy of a BTZ
BH in the presence of gravitational Chern-Simons terms, something
which is difficult to achieve in 3D because there is no manifestly
covariant formulation of the Chern-Simons term, whereas the reduced
theory is manifestly covariant. It is not unlikely that also other
open problems of 3D gravity may be tackled with 2D methods.

\subsection{Integrable systems, Poisson-sigma models  and KdV surfaces}\label{se:psm} 

Some of the pioneering work has been mentioned already in section
\ref{se:FOG} and in table \ref{tab:1}. In two seminal papers by Kummer
and Schwarz \cite{Kummer:1992rt} the usefulness of light-cone gauge
for the Lorentz frame and EF gauge for the curved metric has been
demonstrated for the fourteenth model in table \ref{tab:1}, which is a
rather generic one as it has non-vanishing $U$ and non-monomial $V$. A
Hamiltonian analysis \cite{Grosse:1992vc} revealed an interesting
(W-)algebraic structure of the secondary constraints together with the
fields $X,X^\pm$ as generators. The center of this algebra consists of
the conserved quantity \eqref{eq:c} and its first derivative,
$\partial_1 M$ (which, of course, vanishes on the surface of
constraints).  Consequently, it has been shown by Schaller and Strobl
\cite{Schaller:1994es} that \eqref{eq:FOG} is a special case of a
Poisson-sigma model,\footnote{Dirac-sigma models \cite{Kotov:2004wz}
  are a recent generalization thereof.}
\begin{equation}
  \label{eq:PSM}
  S_{\rm PSM}= - \int_{\mathcal{M}} \left[ X^I \extd A_I - \frac12 P^{IJ} A_J\wedge A_I \right]\,, 
\end{equation}
with a 3D target space, the coordinates of which are
$X^I=\{X,X^+,X^-\}$. The gauge fields comprise the Cartan variables,
$A_I=\{\om,e^-,e^+\}$.  Because the dimension of the Poisson manifold
is odd the Poisson tensor ($I,J\in\{X,\pm\}$)
\begin{equation}
  \label{eq:Ptensor}
  P^{X\pm} =\pm X^\pm\,, \qquad
  P^{+-} = X^+X^-U(X)+V(X)\,, \qquad
  P^{IJ} = -P^{JI}\,,
\end{equation}
cannot have full rank. Therefore, always a Casimir function,
\eqref{eq:c}, exists, which may be interpreted as
``mass''. 
Note that \eqref{eq:Ptensor} indeed fulfills the required
Jacobi-identities, $P^{IL}\partial _{L}P^{JK}+ \mbox{\it perm}\left(
  IJK\right) = 0$.  For a generic (graded) Poisson-sigma model
\eqref{eq:PSM} the commutator of two symmetry transformations
\begin{align}
\label{eq:symtrans}
  \delta X^{I} &= P^{IJ} \varepsilon_{J}\ , & \delta A_{I} &= -\extd \varepsilon_{I}-\left( \partial _{I}P^{JK}\right) \varepsilon_{K}\, A_{J}\ ,
\end{align}
is a (non-linear) symmetry modulo the equations of motion. Only for \(
P^{IJ} \) linear in \( X^{I} \) a Lie algebra is obtained;
cf.~the second model in table \ref{tab:1}. For \eqref{eq:Ptensor} the
symmetries \eqref{eq:symtrans} on-shell correspond to local Lorentz
transformations and diffeomorphisms.  Generalizations discussed in
section \ref{se:top} are particularly transparent in this approach;
essentially, one has to add more target space coordinates to the
Poisson manifold, some of which will be fermionic in supergravity
extensions, cf.~e.g.~\cite{Strobl:1999wv}.

Actually, there exist various approaches to integrability of gravity
models in 2D, cf.~e.g.~\cite{Nicolai:1996pd}, and we can hardly do
them justice here. We will just point out a relation to Korteweg-de
Vries (KdV) surfaces as discussed recently in \cite{Gurses:2005yg}.
These are 2D surfaces embedded in 3D Minkowski space arising from the
KdV equation $\partial_t w=\partial_x^3 w + 6 w \partial_x w$, with
line element (cf.~(11) in \cite{Gurses:2005yg}; $u$ there coincides
with $w$ here) $\extd s^2=2\extd X \extd u-(4\la-w(X,u))\extd u^2$,
where $X\propto x$, $u\propto t$ and $\la$ is some constant. For
static KdV solutions, $\partial_u w=0$, this line element is also a
solution of \eqref{eq:FOG} as can bee seen from \eqref{eq:EF}, with
$\la$ playing the role of the mass $M$. In the non-static case it
describes a solution of \eqref{eq:FOG} coupled to some energy-momentum
tensor. It could be of interest to pursue this relation in more depth.

\subsection{Torsion and non-metricity}

For $U=0$ the equation of motion $R=2V'(X)$, if invertible, allows to
rewrite the action \eqref{eq:GDT} as $S_{\rm R}=\int\extd^2
x\sqrt{-g}f(R)$, cf.~e.g.~\cite{Schmidt:1999wb} and references
therein. As compared to such theories, the literature on models with
torsion $\tau^a={\ast T^a}$,
\begin{equation}
  \label{eq:torsion}
  S_{\rm RT} = \int\extd^2x\sqrt{-g} f(R,\tau^a \tau_a)\,,
\end{equation}
is relatively scarce and consists mainly of elaborations based upon
the fourteenth model in table \ref{tab:1}, where $f=A \tau^a\tau_a + B
R^2 + C R + \La$, also known as ``Poincar{\`e} gauge theory'',
cf.~\cite{Obukhov:1997uc} and references therein. This model in
particular (and a large class of models of type \eqref{eq:torsion})
allows an equivalent reformulation as \eqref{eq:FOG}. Thus, they need
not be discussed separately.

A generalization which includes also effects from non-metricity has
been studied in \cite{Dereli:1994vx}. Elimination of non-metricity
leads again to models of type \eqref{eq:GDT}, \eqref{eq:FOG}, but one
has to be careful with such reformulations as test-particles moving
along geodesics or, alternatively, along auto-parallels, may ``feel''
the difference. Thus, it could be of interest to generalize
\eqref{eq:FOG} (which already contains torsion if $U\neq 0$) as to
include non-metricity, thus dropping the requirement that the
connection $\om^a{}_b$ is proportional to $\eps^a{}_b$. However, a
formulation as Poisson-sigma model \eqref{eq:PSM} (with 6D target
space) seems to be impossible as there are only trivial solutions to
the Jacobi identities.

\subsection{Non-commutative gravity}

In the 1970ies/1980ies theories have been supersymmetrized, in the
1990ies/2000s theories have been ``non-commutativized'', for reviews
cf.~e.g.~\cite{Douglas:2001ba}. The latter procedure
still has not stopped as the original idea, namely to obtain a fully
satisfactory non-commutative version of gravity, has not been achieved
so far. In order to get around the main conceptual obstacles it is
tempting to consider the simplified framework of 2D.

There it is possible to construct non-commutative dilaton gravity
models with a usual (non-twisted) realization of gauge symmetries.%
\footnote{Another approach has been pursued in \cite{Buric:2004rm}.}
A non-commutative version of the Jackiw-Teitelboim model (cf.~the
second entry in table \ref{tab:1}),
\begin{equation}
S_{\rm NCJT}=-\frac{1}{2}\int d^{2}x\,\varepsilon^{\mu\nu}\left[X_{a}\star T_{\mu\nu}^{a}+X_{ab}\star\left(R_{\mu\nu}^{ab}-\Lambda e_{\mu}^{a}\star e_{\nu}^{b}\right)\right]\,,
\label{actJT}
\end{equation}
has been constructed in \cite{Cacciatori:2002ib} and then quantized in
\cite{Vassilevich:2004ym}. 
A non-commutative version of the fourth model in table \ref{tab:1} was
suggested in \cite{Vassilevich:2005fk}.
For a definition of the Moyal-$\star$
and further notation cf.~these two references. A crucial change as
compared to \eqref{eq:FOG}, besides the $\star$, is the appearance of
a second dilaton field $\psi$ in $2X_{ab}= X \eps_{ab}-i\psi
\eta_{ab}$. However, interesting as these results may be, there seems
to be no way to generalize them to generic 2D dilaton gravity without
twisting the gauge symmetries \cite{Vassilevich:2006uv}. Moreover, the
fact that the metric can be changed by ``Lorentz transformations''
seems questionable from a physical point of view,
cf.~\cite{Grumiller:2003df} for a similar problem.

An important step towards constructing a satisfactory non-commutative
gravity was recently made by Wess and collaborators
\cite{Aschieri:2005yw}, who understood how one can construct
diffeomorphism invariants, including the EH action, on non-commutative
spaces (see also \cite{Zupnik:2005ph} for a real formulation). There
is, however, a price to pay. The diffeomorphism group becomes twisted,
i.e., there is a non-trivial coproduct 
\cite{Chaichian:2004za}.  Recently it could be shown
\cite{Vassilevich:2006tc} that twisted gauge
symmetries close for arbitrary gauge groups and thus a construction of
twisted-invariant actions is straightforward. The main element in that
construction (cf.~also
\cite{Chaichian:2004za,Chaichian:2004yh,Aschieri:2005yw,Zupnik:2005ph} and
\cite{Oeckl:2000eg}) is the twist operator
\begin{equation}
\mathcal{F}=\exp \mathcal{P},\qquad
\mathcal{P}= \frac i2 \theta^{\mu\nu} \partial_\mu \otimes \partial_\nu \,,
\label{twist}
\end{equation}
which acts on the tensor products of functions $\phi_1 \otimes \phi_2$.
With the multiplication map 
$\mu (\phi_1 \otimes \phi_2)=\phi_1 \cdot \phi_2$ and
\eqref{twist} the Moyal-Weyl representation
of the star product, 
\begin{equation}
\phi_1 \star \phi_2 = \mu \circ \mathcal{F} (\phi_1 \otimes \phi_2)
= \mu_\star (\phi_1 \otimes \phi_2)\,, 
\label{stapro} 
\end{equation} 
can be constructed. Consider now generators $u$ of some symmetry
transformations which form a Lie algebra. If one knows the action of
these transformations on primary fields, $\delta_u \phi = u \phi$, the
action on tensor products is defined by the coproduct $\Delta$.  In
the undeformed case the coproduct is primitive, $\Delta_0 (u)=u
\otimes 1 + 1 \otimes u$ and $\delta_u (\phi_1 \otimes
\phi_2)=\Delta_0 (u) (\phi_1 \otimes \phi_2) =u\phi_1 \otimes \phi_2 +
\phi_1 \otimes u\phi_2$ satisfies the usual Leibniz rule. The action
of symmetry generators on elementary fields is left undeformed, but
the coproduct is twisted,
\begin{equation}
\Delta (u) = \exp (-\mathcal{P}) \Delta_0(u) \exp ( \mathcal{P})\,.
\label{defDel}
\end{equation}
Obviously, twisting preserves the commutation relations. Therefore,
the commutators of gauge transformations for an arbitrary gauge group close.

It seems plausible that a corresponding generalization to twisted
non-linear gauge symmetries will be a crucial technical pre-requisite
to a successful construction of generic non-commutative 2D dilaton
gravity.\footnote{The relation of \eqref{eq:PSM} to a specific Lie
  algebroid \cite{Bojowald:2004wu} could be helpful in this context.}
It would allow, among other things, a thorough discussion of
non-commutative BHs, along the lines of sections
\ref{se:1}-\ref{se:4}.

\acknowledgments{ DG and RM would like to thank cordially L.~Bergamin,
  W.~Kummer and D.~Vassilevich for a long-time collaboration and
  helpful discussions, respectively, on most of the topics reviewed in
  this work.  Moreover, DG is grateful to M.~Adak, P.~Aichelburg,
  S.~Alexandrov, H.~Balasin, M.~Bojowald, M.~Cadoni, S.~Carlip,
  T.~Dereli, M.~G\"urses, A.~Iorio, R.~Jackiw, M.~Katanaev,
  C.~Lechner, F.~Meyer, S.~Mignemi, C.~Nu{\~n}ez, Y.~Obukhov,
  M.-I.~Park, M.~P\"urrer, R.~Sch\"utzhold, T.~Strobl, W.~Unruh,
  P.~van Nieuwenhuizen and S.~Weinfurtner for helpful discussions
  and/or correspondence.  In addition, DG would like to thank the
  organizers of the Fifth Workshop on QUANTIZATION, DUALITIES AND
  INTEGRABLE SYSTEMS in Denizli, Turkey, in particular M.~Adak for the
  kind
  invitation. 

  DG has been supported by project GR-3157/1-1 of the German Research
  Foundation (DFG). Additional financial support due to Pamukkale
  University is acknowledged gratefully.  RM has been supported
  financially by the MPI and expresses his gratitude to J.~Jost in
  this regard.  }



\begin{thebibliography}{10%
0}

\bibitem{Square:1884}
Such a life, with all vision limited to a Point, and all motion to a Straight
  Line, seemed to me inexpressibly dreary; and I was surprised to note the
  vivacity and cheerfulness of the King. [Edwin A.~Abbot, ``Flatland --- A
  Romance of Many Dimensions.'' Dover Publications 1992, New York. (first
  published under the pseudonym A. Square in 1884, Seeley \& Co., London)].

\bibitem{Brown:1988}
J.~Brown, {\em Lower Dimensional Gravity}.
\newblock World Scientific, 1988.

\bibitem{Polyakov:1987zb}
A.~M. Polyakov, ``Quantum gravity in two-dimensions,'' {\em Mod. Phys. Lett.}
  {\bf A2} (1987)
893.

\bibitem{Grumiller:2002nm}
D.~Grumiller, W.~Kummer, and D.~V. Vassilevich, ``Dilaton gravity in two
  dimensions,'' {\em Phys. Rept.} {\bf 369} (2002) 327--429,
\href{http://arXiv.org/abs/hep-th/0204253}{{\tt hep-th/0204253}}.

\bibitem{Thomi:1984na}
P.~Thomi, B.~Isaak, and P.~H{\'a}j{\'i}{\v{c}}ek, ``Spherically symmetric
  systems of fields and black holes. 1. {D}efinition and properties of apparent
  horizon,'' {\em Phys. Rev.} {\bf D30} (1984)
1168.

P.~H{\'a}j{\'i}{\v{c}}ek, ``Spherically symmetric systems of fields and black
  holes. 2. {A}pparent horizon in canonical formalism,'' {\em Phys. Rev.} {\bf
  D30} (1984)
1178.

\bibitem{Teitelboim:1983ux}
C.~Teitelboim, ``Gravitation and {H}amiltonian structure in two space-time
  dimensions,'' {\em Phys. Lett.} {\bf B126} (1983)
41.

\bibitem{Jackiw:1985je}
R.~Jackiw, ``Lower dimensional gravity,'' {\em Nucl. Phys.} {\bf B252} (1985)
343--356.

\bibitem{Witten:1991yr}
E.~Witten, ``On string theory and black holes,'' {\em Phys. Rev.} {\bf D44}
  (1991)
314--324.

G.~Mandal, A.~M. Sengupta, and S.~R. Wadia, ``Classical solutions of
  two-dimensional string theory,'' {\em Mod. Phys. Lett.} {\bf A6} (1991)
1685--1692.

S.~Elitzur, A.~Forge, and E.~Rabinovici, ``Some global aspects of string
  compactifications,'' {\em Nucl. Phys.} {\bf B359} (1991)
581--610.

\bibitem{Callan:1992rs}
C.~G. Callan, Jr., S.~B. Giddings, J.~A. Harvey, and A.~Strominger,
  ``Evanescent black holes,'' {\em Phys. Rev.} {\bf D45} (1992) 1005--1009,
\href{http://www.arXiv.org/abs/hep-th/9111056}{{\tt hep-th/9111056}}.

\bibitem{Lemos:1994py}
J.~P.~S. Lemos and P.~M. Sa, ``The black holes of a general two-dimensional
  dilaton gravity theory,'' {\em Phys. Rev.} {\bf D49} (1994) 2897--2908,
\href{http://www.arXiv.org/abs/arXiv:gr-qc/9311008}{{\tt arXiv:gr-qc/9311008}}.

\bibitem{Fabbri:1996bz}
A.~Fabbri and J.~G. Russo, ``Soluble models in 2d dilaton gravity,'' {\em Phys.
  Rev.} {\bf D53} (1996) 6995--7002,
\href{http://arXiv.org/abs/hep-th/9510109}{{\tt hep-th/9510109}}.

\bibitem{Grumiller:2003hq}
D.~Grumiller, ``{Long time black hole evaporation with bounded Hawking flux},''
  {\em JCAP} {\bf 05} (2004) 005,
\href{http://www.arXiv.org/abs/gr-qc/0307005}{{\tt gr-qc/0307005}}.

\bibitem{Katanaev:1997ni}
M.~O. Katanaev, W.~Kummer, and H.~Liebl, ``On the completeness of the black
  hole singularity in 2d dilaton theories,'' {\em Nucl. Phys.} {\bf B486}
  (1997) 353--370,
\href{http://www.arXiv.org/abs/gr-qc/9602040}{{\tt gr-qc/9602040}}.

\bibitem{Nakayama:2004vk}
Y.~Nakayama, ``{Liouville field theory: A decade after the revolution},'' {\em
  Int. J. Mod. Phys.} {\bf A19} (2004) 2771--2930,
\href{http://www.arXiv.org/abs/hep-th/0402009}{{\tt hep-th/0402009}}.

\bibitem{Grumiller:2002dm}
D.~Grumiller, W.~Kummer, and D.~V. Vassilevich, ``Virtual black holes in
  generalized dilaton theories (and their special role in string gravity),''
  {\em European Phys. J.} {\bf C30} (2003) 135--143,
\href{http://arXiv.org/abs/hep-th/0208052}{{\tt hep-th/0208052}}.

\bibitem{Reissner:1916}
H.~Reissner, ``{\"U}ber die {E}igengravitation des elektrischen {F}eldes nach
  der {E}insteinschen {T}heorie,'' {\em Ann. Phys.} {\bf 50} (1916) 106.

G.~Nordstr{\"o}m, ``On the energy of the gravitation field in {E}instein's
  theory,'' {\em Proc. Kon. Ned. Akad. Wet.} {\bf 20} (1916) 1238.

\bibitem{Hawking:1982dh}
S.~W. Hawking and D.~N. Page, ``Thermodynamics of black holes in anti-de
  {S}itter space,'' {\em Commun. Math. Phys.} {\bf 87} (1983)
577.

\bibitem{Katanaev:1986wk}
M.~O. Katanaev and I.~V. Volovich, ``String model with dynamical geometry and
  torsion,'' {\em Phys. Lett.} {\bf B175} (1986)
413--416;
``Two-dimensional gravity with dynamical
  torsion and strings,'' {\em Ann. Phys.} {\bf 197} (1990) 1.

\bibitem{Achucarro:1993fd}
A.~Achucarro and M.~E. Ortiz, ``Relating black holes in two-dimensions and
  three- dimensions,'' {\em Phys. Rev.} {\bf D48} (1993) 3600--3605,
\href{http://www.arXiv.org/abs/hep-th/9304068}{{\tt hep-th/9304068}}.

\bibitem{Guralnik:2003we}
G.~Guralnik, A.~Iorio, R.~Jackiw, and S.~Y. Pi, ``{Dimensionally reduced
  gravitational Chern-Simons term and its kink},'' {\em Ann. Phys.} {\bf 308}
  (2003) 222--236,
\href{http://www.arXiv.org/abs/hep-th/0305117}{{\tt hep-th/0305117}}.

\bibitem{Grumiller:2003ad}
D.~Grumiller and W.~Kummer, ``{The classical solutions of the dimensionally
  reduced gravitational Chern-Simons theory},'' {\em Ann. Phys.} {\bf 308}
  (2003) 211--221,
\href{http://www.arXiv.org/abs/hep-th/0306036}{{\tt hep-th/0306036}}.

L.~Bergamin, D.~Grumiller, A.~Iorio, and C.~Nu{\~n}ez, ``{Chemistry of
  Chern-Simons supergravity: Reduction to a BPS kink, oxidation to M-theory and
  thermodynamical aspects},'' {\em JHEP} {\bf 11} (2004) 021,
\href{http://www.arXiv.org/abs/hep-th/0409273}{{\tt hep-th/0409273}}.

\bibitem{Bergamin:2005au}
L.~Bergamin, ``Constant dilaton vacua and kinks in 2d (super-)gravity,''
\href{http://www.arXiv.org/abs/hep-th/0509183}{{\tt hep-th/0509183}}.

\bibitem{Douglas:2003up}
M.~R. Douglas {\em et al.}, ``A new hat for the c = 1 matrix model,''
\href{http://www.arXiv.org/abs/hep-th/0307195}{{\tt hep-th/0307195}}.

\bibitem{Gukov:2003yp}
S.~Gukov, T.~Takayanagi, and N.~Toumbas, ``{Flux backgrounds in 2D string
  theory},'' {\em JHEP} {\bf 03} (2004) 017,
\href{http://www.arXiv.org/abs/hep-th/0312208}{{\tt hep-th/0312208}}.

\bibitem{Dijkgraaf:1992ba}
R.~Dijkgraaf, H.~Verlinde, and E.~Verlinde, ``String propagation in a black
  hole geometry,'' {\em Nucl. Phys.} {\bf B371} (1992)
269--314.

\bibitem{Grumiller:2005sq}
D.~Grumiller, ``An action for the exact string black hole,'' {\em JHEP} {\bf
  05} (2005) 028,
\href{http://www.arXiv.org/abs/hep-th/0501208}{{\tt hep-th/0501208}}.

\bibitem{Isler:1989hq}
K.~Isler and C.~A. Trugenberger, ``A gauge theory of two-dimensional quantum
  gravity,'' {\em Phys. Rev. Lett.} {\bf 63} (1989)
834.

A.~H. Chamseddine and D.~Wyler, ``Gauge theory of topological gravity in
  (1+1)-dimensions,'' {\em Phys. Lett.} {\bf B228} (1989)
75.

\bibitem{Verlinde:1991rf}
H.~Verlinde, ``Black holes and strings in two dimensions,'' in {\em Trieste
  Spring School on Strings and Quantum Gravity}, pp.~178--207.
\newblock April, 1991.
\newblock the same lectures have been given at MGVI in Japan, June, 1991.

\bibitem{Cangemi:1992bj}
D.~Cangemi and R.~Jackiw, ``Gauge invariant formulations of lineal gravity,''
  {\em Phys. Rev. Lett.} {\bf 69} (1992) 233--236,
\href{http://arXiv.org/abs/hep-th/9203056}{{\tt hep-th/9203056}}.

A.~Achucarro, ``Lineal gravity from planar gravity,'' {\em Phys. Rev. Lett.}
  {\bf 70} (1993) 1037--1040,
\href{http://www.arXiv.org/abs/hep-th/9207108}{{\tt hep-th/9207108}}.

\bibitem{Ikeda:1993aj}
N.~Ikeda and K.~I. Izawa, ``General form of dilaton gravity and nonlinear gauge
  theory,'' {\em Prog. Theor. Phys.} {\bf 90} (1993) 237--246,
\href{http://www.arXiv.org/abs/hep-th/9304012}{{\tt hep-th/9304012}}.

\bibitem{Schaller:1994es}
P.~Schaller and T.~Strobl, ``Poisson structure induced (topological) field
  theories,'' {\em Mod. Phys. Lett.} {\bf A9} (1994) 3129--3136,
\href{http://arXiv.org/abs/hep-th/9405110}{{\tt hep-th/9405110}}.

\bibitem{Russo:1992yg}
J.~G. Russo and A.~A. Tseytlin, ``Scalar tensor quantum gravity in
  two-dimensions,'' {\em Nucl. Phys.} {\bf B382} (1992) 259--275,
\href{http://www.arXiv.org/abs/arXiv:hep-th/9201021}{{\tt
  arXiv:hep-th/9201021}}.

S.~D. Odintsov and I.~L. Shapiro, ``One loop renormalization of two-dimensional
  induced quantum gravity,'' {\em Phys. Lett.} {\bf B263} (1991)
183--189.

T.~Banks and M.~O'Loughlin, ``Two-dimensional quantum gravity in {M}inkowski
  space,'' {\em Nucl. Phys.} {\bf B362} (1991)
649--664.

R.~B. Mann, A.~Shiekh, and L.~Tarasov, ``Classical and quantum properties of
  two-dimensional black holes,'' {\em Nucl. Phys.} {\bf B341} (1990)
134--154.

\bibitem{Klosch:1996qv}
T.~Kl{\"o}sch and T.~Strobl, ``Classical and quantum gravity in 1+1 dimensions.
  {P}art {II}: {T}he universal coverings,'' {\em Class. Quant. Grav.} {\bf 13}
  (1996) 2395--2422,
\href{http://www.arXiv.org/abs/arXiv:gr-qc/9511081}{{\tt arXiv:gr-qc/9511081}}.

\bibitem{Klosch:1996fi}
T.~Kl{\"o}sch and T.~Strobl, ``Classical and quantum gravity in
  (1+1)-dimensions. {P}art {I}: {A} unifying approach,'' {\em Class. Quant.
  Grav.} {\bf 13} (1996) 965--984,
\href{http://www.arXiv.org/abs/arXiv:gr-qc/9508020}{{\tt arXiv:gr-qc/9508020}}.

\bibitem{Hao:2003aa}
J.-G. Hao and X.-Z. Li, ``{Constructing dark energy models with late time de
  Sitter attractor},'' {\em Phys. Rev.} {\bf D68} (2003) 083514,
\href{http://www.arXiv.org/abs/hep-th/0306033}{{\tt hep-th/0306033}}.

\bibitem{Bergamin:2003mh}
L.~Bergamin, D.~Grumiller, and W.~Kummer, ``Supersymmetric black holes in 2d
  dilaton supergravity: baldness and extremality,'' {\em J. Phys.} {\bf A37}
  (2004) 3881--3901,
\href{http://www.arXiv.org/abs/hep-th/0310006}{{\tt hep-th/0310006}}.

\bibitem{Thompson:2003fz}
D.~M. Thompson, ``{AdS solutions of 2D type 0A},'' {\em Phys. Rev.} {\bf D70}
  (2004) 106001,
\href{http://www.arXiv.org/abs/hep-th/0312156}{{\tt hep-th/0312156}}.

\bibitem{Birmingham:1991ty}
D.~Birmingham, M.~Blau, M.~Rakowski, and G.~Thompson, ``Topological field
  theory,'' {\em Phys. Rept.} {\bf 209} (1991)
129--340.

\bibitem{Strobl:1999wv}
T.~Strobl, ``Gravity in two spacetime dimensions,''
  \href{http://www.arXiv.org/abs/hep-th/0011240}{{\tt hep-th/0011240}}.
Habilitation thesis.

\bibitem{Park:1993sd}
Y.-C. Park and A.~Strominger, ``Supersymmetry and positive energy in classical
  and quantum two-dimensional dilaton gravity,'' {\em Phys. Rev.} {\bf D47}
  (1993) 1569--1575,
\href{http://www.arXiv.org/abs/arXiv:hep-th/9210017}{{\tt
  arXiv:hep-th/9210017}}.

J.~M. Izquierdo, ``Free differential algebras and generic 2d dilatonic
  (super)gravities,'' {\em Phys. Rev.} {\bf D59} (1999) 084017,
\href{http://www.arXiv.org/abs/arXiv:hep-th/9807007}{{\tt
  arXiv:hep-th/9807007}}.

T.~Strobl, ``Target-superspace in 2d dilatonic supergravity,'' {\em Phys.
  Lett.} {\bf B460} (1999) 87--93,
\href{http://www.arXiv.org/abs/arXiv:hep-th/9906230}{{\tt
  arXiv:hep-th/9906230}}.

M.~Ertl, W.~Kummer, and T.~Strobl, ``General two-dimensional supergravity from
  {P}oisson superalgebras,'' {\em JHEP} {\bf 01} (2001) 042,
\href{http://www.arXiv.org/abs/arXiv:hep-th/0012219}{{\tt
  arXiv:hep-th/0012219}}.

M.~Ertl, {\em Supergravity in two spacetime dimensions}.
\newblock PhD thesis, {T}echnische {U}niversit{\"a}t {W}ien, 2001.
\newblock
\href{http://www.arXiv.org/abs/arXiv:hep-th/0102140}{{\tt
  arXiv:hep-th/0102140}}.
\newblock

L.~Bergamin and W.~Kummer, ``{The complete solution of 2D superfield
  supergravity from graded Poisson-Sigma models and the super pointparticle},''
  {\em Phys. Rev.} {\bf D68} (2003) 104005,
\href{http://www.arXiv.org/abs/hep-th/0306217}{{\tt hep-th/0306217}};
``Two-dimensional {N}=(2,2) dilaton supergravity
  from graded {P}oisson-{S}igma models {I}: Complete actions and their
  symmetries.,'' {\em Eur. Phys. J.} {\bf C39} (2005) S41--S52,
\href{http://www.arXiv.org/abs/hep-th/0402138}{{\tt hep-th/0402138}};
``Two-dimensional {N} = (2,2) dilaton supergravity
  from graded {P}oisson-{S}igma models. {II}: Analytic solution and {BPS}
  states,'' {\em Eur. Phys. J.} {\bf C39} (2005) S53--S63,
\href{http://www.arXiv.org/abs/hep-th/0411204}{{\tt hep-th/0411204}}.

L.~Bergamin, D.~Grumiller, and W.~Kummer, ``Quantization of 2d dilaton
  supergravity with matter,'' {\em JHEP} {\bf 05} (2004) 060,
\href{http://www.arXiv.org/abs/hep-th/0404004}{{\tt hep-th/0404004}}.

\bibitem{Bergamin:2002ju}
L.~Bergamin and W.~Kummer, ``{Graded Poisson sigma models and dilaton-deformed
  2d supergravity algebra},'' {\em JHEP} {\bf 05} (2003) 074,
\href{http://www.arXiv.org/abs/hep-th/0209209}{{\tt hep-th/0209209}}.

\bibitem{Meyer:2005fz}
R.~Meyer, ``Constraints in two-dimensional dilaton gravity with fermions,''
\href{http://www.arXiv.org/abs/hep-th/0512267}{{\tt hep-th/0512267}}.

\bibitem{Thirring:1958in}
W.~E. Thirring, ``A soluble relativistic field theory,'' {\em Annals Phys.}
  {\bf 3} (1958)
91--112.

\bibitem{Balasin:2004gf}
H.~Balasin, C.~G. Boehmer, and D.~Grumiller, ``The spherically symmetric
  standard model with gravity,'' {\em Gen. Rel. Grav.} {\bf 37} (2005)
  1435--1482,
\href{http://www.arXiv.org/abs/gr-qc/0412098}{{\tt gr-qc/0412098}}.

\bibitem{Kummer:1992ef}
W.~Kummer, ``Deformed {ISO}(2,1) symmetry and non-{E}insteinian 2d-gravity with
  matter,'' in {\em HADRON STRUCTURE '92}, D.~Bruncko and J.~Urban, eds.
\newblock September, 1992.
\newblock Stara Lesna, Czechoslovakia.

\bibitem{Pelzer:1998ea}
H.~Pelzer and T.~Strobl, ``Generalized 2d dilaton gravity with matter fields,''
  {\em Class. Quant. Grav.} {\bf 15} (1998) 3803--3825,
\href{http://www.arXiv.org/abs/arXiv:gr-qc/9805059}{{\tt arXiv:gr-qc/9805059}}.

\bibitem{Aichelburg:1971dh}
P.~C. Aichelburg and R.~U. Sexl, ``On the gravitational field of a massless
  particle,'' {\em Gen. Rel. Grav.} {\bf 2} (1971)
303--312.

\bibitem{Balasin:2003cn}
H.~Balasin and D.~Grumiller, ``The ultrarelativistic limit of 2d dilaton
  gravity and its energy momentum tensor,'' {\em Class. Quant. Grav.} {\bf 21}
  (2004) 2859--2872,
\href{http://www.arXiv.org/abs/gr-qc/0312086}{{\tt gr-qc/0312086}}.

\bibitem{Fisher:1948yn}
I.~Z. Fisher, ``Scalar mesostatic field with regard for gravitational
  effects,'' {\em Zh. Eksp. Teor. Fiz.} {\bf 18} (1948) 636--640,
\href{http://arXiv.org/abs/gr-qc/9911008}{{\tt gr-qc/9911008}}.

\bibitem{Filippov:2002sp}
A.~T. Filippov and D.~Maison, ``Horizons in 1+1 dimensional dilaton gravity
  coupled to matter,'' {\em Class. Quant. Grav.} {\bf 20} (2003) 1779--1786,
\href{http://www.arXiv.org/abs/gr-qc/0210081}{{\tt gr-qc/0210081}}.

\bibitem{Grumiller:2004wi}
D.~Grumiller and D.~Mayerhofer, ``On static solutions in 2d dilaton gravity
  with scalar matter,'' {\em Class. Quant. Grav.} {\bf 21} (2004) 5893--5914,
\href{http://www.arXiv.org/abs/gr-qc/0404013}{{\tt gr-qc/0404013}}.

\bibitem{Wyman:1981bd}
M.~Wyman, ``Static spherically symmetric scalar fields in general relativity,''
  {\em Phys. Rev.} {\bf D24} (1981)
839--841.

\bibitem{Bilge:2005vx}
A.~H. Bilge and D.~Daghan, ``{Partial decoupling and exact static solutions for
  Choptuik's spacetime},''
\href{http://www.arXiv.org/abs/gr-qc/0508020}{{\tt gr-qc/0508020}}.

\bibitem{Roberts:1989sk}
M.~D. Roberts, ``Scalar field counterexamples to the cosmic censorship
  hypothesis,'' {\em Gen. Rel. Grav.} {\bf 21} (1989)
907--939.

\bibitem{Takayanagi:2003sm}
T.~Takayanagi and N.~Toumbas, ``{A matrix model dual of type 0B string theory
  in two dimensions},'' {\em JHEP} {\bf 07} (2003) 064,
\href{http://www.arXiv.org/abs/hep-th/0307083}{{\tt hep-th/0307083}}.

\bibitem{Ginsparg:1993is}
P.~Ginsparg and G.~W. Moore, ``Lectures on 2-d gravity and 2-d string theory,''
\href{http://arXiv.org/abs/hep-th/9304011}{{\tt hep-th/9304011}}.

P.~Di~Francesco, P.~H. Ginsparg, and J.~Zinn-Justin, ``{2-D Gravity and random
  matrices},'' {\em Phys. Rept.} {\bf 254} (1995) 1--133,
\href{http://www.arXiv.org/abs/hep-th/9306153}{{\tt hep-th/9306153}}.

S.~Alexandrov, ``Matrix quantum mechanics and two-dimensional string theory in
  non-trivial backgrounds,''
\href{http://www.arXiv.org/abs/hep-th/0311273}{{\tt hep-th/0311273}}.

\bibitem{Kazakov:2000pm}
V.~Kazakov, I.~K. Kostov, and D.~Kutasov, ``A matrix model for the
  two-dimensional black hole,'' {\em Nucl. Phys.} {\bf B622} (2002) 141--188,
\href{http://www.arXiv.org/abs/hep-th/0101011}{{\tt hep-th/0101011}}.

\bibitem{Bergamin:2004pn}
L.~Bergamin, D.~Grumiller, W.~Kummer, and D.~V. Vassilevich, ``{Classical and
  quantum integrability of 2D dilaton gravities in Euclidean space},'' {\em
  Class. Quant. Grav.} {\bf 22} (2005) 1361--1382,
\href{http://www.arXiv.org/abs/hep-th/0412007}{{\tt hep-th/0412007}}.

\bibitem{Klebanov:1998yy}
I.~R. Klebanov and A.~A. Tseytlin, ``D-branes and dual gauge theories in type 0
  strings,'' {\em Nucl. Phys.} {\bf B546} (1999) 155--181,
\href{http://www.arXiv.org/abs/hep-th/9811035}{{\tt hep-th/9811035}}.

A.~Strominger, ``{A matrix model for AdS(2)},'' {\em JHEP} {\bf 03} (2004) 066,
\href{http://www.arXiv.org/abs/hep-th/0312194}{{\tt hep-th/0312194}}.

J.~L. Davis, L.~A. Pando~Zayas, and D.~Vaman, ``{On black hole thermodynamics
  of 2-D type 0A},'' {\em JHEP} {\bf 03} (2004) 007,
\href{http://www.arXiv.org/abs/hep-th/0402152}{{\tt hep-th/0402152}}.

U.~H. Danielsson, J.~P. Gregory, M.~E. Olsson, P.~Rajan, and M.~Vonk, ``{Type
  0A 2D black hole thermodynamics and the deformed matrix model},'' {\em JHEP}
  {\bf 04} (2004) 065,
\href{http://www.arXiv.org/abs/hep-th/0402192}{{\tt hep-th/0402192}}.

J.~L. Davis and R.~McNees, ``{Boundary counterterms and the thermodynamics of
  2-D black holes},'' {\em JHEP} {\bf 09} (2005) 072,
\href{http://www.arXiv.org/abs/hep-th/0411121}{{\tt hep-th/0411121}}.

\bibitem{Tseytlin:1991ht}
A.~A. Tseytlin, ``On the form of the black hole solution in d = 2 theory,''
  {\em Phys. Lett.} {\bf B268} (1991)
175--178.

I.~Jack, D.~R.~T. Jones, and J.~Panvel, ``Exact bosonic and supersymmetric
  string black hole solutions,'' {\em Nucl. Phys.} {\bf B393} (1993) 95--110,
\href{http://arXiv.org/abs/hep-th/9201039}{{\tt hep-th/9201039}}.

\bibitem{Tseytlin:1992ri}
A.~A. Tseytlin, ``{Effective action of gauged WZW model and exact string
  solutions},'' {\em Nucl. Phys.} {\bf B399} (1993) 601--622,
\href{http://www.arXiv.org/abs/hep-th/9301015}{{\tt hep-th/9301015}}.

I.~Bars and K.~Sfetsos, ``{Exact effective action and space-time geometry in
  gauged WZW models},'' {\em Phys. Rev.} {\bf D48} (1993) 844--852,
\href{http://www.arXiv.org/abs/hep-th/9301047}{{\tt hep-th/9301047}}.

\bibitem{Becker:1994vd}
K.~Becker, ``Strings, black holes and conformal field theory,''
\href{http://arXiv.org/abs/hep-th/9404157}{{\tt hep-th/9404157}}.

\bibitem{Kazakov:2001pj}
V.~A. Kazakov and A.~A. Tseytlin, ``On free energy of 2-d black hole in bosonic
  string theory,'' {\em JHEP} {\bf 06} (2001) 021,
\href{http://arXiv.org/abs/hep-th/0104138}{{\tt hep-th/0104138}}.

\bibitem{Perry:1993ry}
M.~J. Perry and E.~Teo, ``Nonsingularity of the exact two-dimensional string
  black hole,'' {\em Phys. Rev. Lett.} {\bf 70} (1993) 2669--2672,
\href{http://arXiv.org/abs/hep-th/9302037}{{\tt hep-th/9302037}}.

P.~Yi, ``Nonsingular {2-D} black holes and classical string backgrounds,'' {\em
  Phys. Rev.} {\bf D48} (1993) 2777--2788,
\href{http://arXiv.org/abs/hep-th/9302070}{{\tt hep-th/9302070}}.

\bibitem{Grumiller:2002md}
D.~Grumiller and D.~V. Vassilevich, ``Non-existence of a dilaton gravity action
  for the exact string black hole,'' {\em JHEP} {\bf 11} (2002) 018,
\href{http://arXiv.org/abs/hep-th/0210060}{{\tt hep-th/0210060}}.

\bibitem{Kummer:1995qv}
W.~Kummer and P.~Widerin, ``Conserved quasilocal quantities and general
  covariant theories in two-dimensions,'' {\em Phys. Rev.} {\bf D52} (1995)
  6965--6975,
\href{http://www.arXiv.org/abs/arXiv:gr-qc/9502031}{{\tt arXiv:gr-qc/9502031}}.

\bibitem{Frolov:1998}
V.~Frolov and I.~Novikov, {\em {Black Hole Physics}}.
\newblock Kluwer Academic Publishers, 1998.

\bibitem{Faddeev:1982id}
L.~D. Faddeev, ``The energy problem in {E}instein's theory of gravitation,''
  {\em Sov. Phys. Usp.} {\bf 25} (1982)
130--142.

\bibitem{Liebl:1997ti}
H.~Liebl, D.~V. Vassilevich, and S.~Alexandrov, ``Hawking radiation and masses
  in generalized dilaton theories,'' {\em Class. Quant. Grav.} {\bf 14} (1997)
  889--904,
\href{http://www.arXiv.org/abs/arXiv:gr-qc/9605044}{{\tt arXiv:gr-qc/9605044}}.

\bibitem{Frolov:1992xx}
V.~P. Frolov, ``Two-dimensional black hole physics,'' {\em Phys. Rev.} {\bf
  D46} (1992)
5383--5394.

\bibitem{Mann:1993yv}
R.~B. Mann, ``Conservation laws and 2-d black holes in dilaton gravity,'' {\em
  Phys. Rev.} {\bf D47} (1993) 4438--4442,
\href{http://www.arXiv.org/abs/hep-th/9206044}{{\tt hep-th/9206044}}.

\bibitem{Grosse:1992vc}
H.~Grosse, W.~Kummer, P.~Presnajder, and D.~J. Schwarz, ``{Novel symmetry of
  nonEinsteinian gravity in two- dimensions},'' {\em J. Math. Phys.} {\bf 33}
  (1992) 3892--3900,
\href{http://www.arXiv.org/abs/hep-th/9205071}{{\tt hep-th/9205071}}.

\bibitem{Wald:1999vt}
R.~M. Wald, ``The thermodynamics of black holes,'' {\em Living Rev. Rel.} {\bf
  4} (2001) 6,
\href{http://www.arXiv.org/abs/gr-qc/9912119}{{\tt gr-qc/9912119}}.

\bibitem{Gibbons:1994cg}
G.~W. Gibbons and S.~W. Hawking, eds., {\em Euclidean quantum gravity}.
\newblock Singapore: World Scientific, 1993.

\bibitem{Kummer:1999zy}
W.~Kummer and D.~V. Vassilevich, ``{Hawking radiation from dilaton gravity in
  (1+1) dimensions: A pedagogical review},'' {\em Annalen Phys.} {\bf 8} (1999)
  801--827,
\href{http://arXiv.org/abs/gr-qc/9907041}{{\tt gr-qc/9907041}}.

\bibitem{Wald:1993nt}
R.~M. Wald, ``Black hole entropy is the {N}{\"o}ther charge,'' {\em Phys. Rev.}
  {\bf D48} (1993) 3427--3431,
\href{http://arXiv.org/abs/gr-qc/9307038}{{\tt gr-qc/9307038}}.

V.~Iyer and R.~M. Wald, ``Some properties of {N}{\"o}ther charge and a proposal
  for dynamical black hole entropy,'' {\em Phys. Rev.} {\bf D50} (1994)
  846--864,
\href{http://arXiv.org/abs/gr-qc/9403028}{{\tt gr-qc/9403028}}.

\bibitem{Gegenberg:1995pv}
J.~Gegenberg, G.~Kunstatter, and D.~Louis-Martinez, ``Observables for
  two-dimensional black holes,'' {\em Phys. Rev.} {\bf D51} (1995) 1781--1786,
\href{http://arXiv.org/abs/gr-qc/9408015}{{\tt gr-qc/9408015}}.

\bibitem{Bloete:1986qm}
H.~W.~J. Bloete, J.~L. Cardy, and M.~P. Nightingale, ``Conformal invariance,
  the central charge, and universal finite size amplitudes at criticality,''
  {\em Phys. Rev. Lett.} {\bf 56} (1986)
742--745.

J.~L. Cardy, ``Operator content of two-dimensional conformally invariant
  theories,'' {\em Nucl. Phys.} {\bf B270} (1986)
186--204.

\bibitem{Strominger:1996sh}
A.~Strominger and C.~Vafa, ``{Microscopic Origin of the Bekenstein-Hawking
  Entropy},'' {\em Phys. Lett.} {\bf B379} (1996) 99--104,
\href{http://www.arXiv.org/abs/hep-th/9601029}{{\tt hep-th/9601029}}.

K.~V. Krasnov, ``On statistical mechanics of gravitational systems,'' {\em Gen.
  Rel. Grav.} {\bf 30} (1998) 53--68,
\href{http://www.arXiv.org/abs/gr-qc/9605047}{{\tt gr-qc/9605047}}.

A.~Ashtekar, J.~Baez, A.~Corichi, and K.~Krasnov, ``Quantum geometry and black
  hole entropy,'' {\em Phys. Rev. Lett.} {\bf 80} (1998) 904--907,
\href{http://www.arXiv.org/abs/gr-qc/9710007}{{\tt gr-qc/9710007}}.

M.~Cadoni and S.~Mignemi, ``Entropy of 2d black holes from counting
  microstates,'' {\em Phys. Rev.} {\bf D59} (1999) 081501,
\href{http://www.arXiv.org/abs/hep-th/9810251}{{\tt hep-th/9810251}}.

S.~Carlip, ``Black hole entropy from conformal field theory in any dimension,''
  {\em Phys. Rev. Lett.} {\bf 82} (1999) 2828--2831,
\href{http://www.arXiv.org/abs/hep-th/9812013}{{\tt hep-th/9812013}}.

S.~N. Solodukhin, ``Conformal description of horizon's states,'' {\em Phys.
  Lett.} {\bf B454} (1999) 213--222,
\href{http://www.arXiv.org/abs/hep-th/9812056}{{\tt hep-th/9812056}}.

M.-I. Park and J.~H. Yee, ``{Comments on 'Entropy of 2D black holes from
  counting microstates'},'' {\em Phys. Rev.} {\bf D61} (2000) 088501,
\href{http://www.arXiv.org/abs/hep-th/9910213}{{\tt hep-th/9910213}}.

S.~Carlip, ``{Reply to the comment by Park and Ho on 'Black hole entropy from
  conformal field theory in any dimension'},'' {\em Phys. Rev. Lett.} {\bf 83}
  (1999) 5596,
\href{http://www.arXiv.org/abs/hep-th/9910247}{{\tt hep-th/9910247}}.

M.-I. Park and J.~Ho, ``{Comments on 'Black hole entropy from conformal field
  theory in any dimension'},'' {\em Phys. Rev. Lett.} {\bf 83} (1999) 5595,
\href{http://www.arXiv.org/abs/hep-th/9910158}{{\tt hep-th/9910158}}.

I.~Sachs and S.~N. Solodukhin, ``Horizon holography,'' {\em Phys. Rev.} {\bf
  D64} (2001) 124023,
\href{http://arXiv.org/abs/hep-th/0107173}{{\tt hep-th/0107173}}.

M.-I. Park, ``{Hamiltonian dynamics of bounded spacetime and black hole
  entropy: Canonical method},'' {\em Nucl. Phys.} {\bf B634} (2002) 339--369,
\href{http://www.arXiv.org/abs/hep-th/0111224}{{\tt hep-th/0111224}}.

N.~Pinamonti and L.~Vanzo, ``Central charges and boundary fields for two
  dimensional dilatonic black holes,'' {\em Phys. Rev.} {\bf D69} (2004)
  084012,
\href{http://www.arXiv.org/abs/hep-th/0312065}{{\tt hep-th/0312065}}.

G.~Kang, J.-I. Koga, and M.-I. Park, ``Near-horizon conformal symmetry and
  black hole entropy in any dimension,'' {\em Phys. Rev.} {\bf D70} (2004)
  024005,
\href{http://www.arXiv.org/abs/hep-th/0402113}{{\tt hep-th/0402113}}.

\bibitem{Grumiller:2005vy}
D.~Grumiller, ``Logarithmic corrections to the entropy of the exact string
  black hole,'' in {\em Path Integrals from Quantum Information to Cosmology},
  C.~Burdik, N.~Navratil, and S.~Posta, eds.
\newblock JINR Publishing Department, Prague, June, 2005.
\newblock
\href{http://www.arXiv.org/abs/hep-th/0506175}{{\tt hep-th/0506175}}.
\newblock

\bibitem{Aharony:1999ti}
O.~Aharony, S.~S. Gubser, J.~M. Maldacena, H.~Ooguri, and Y.~Oz, ``{Large N
  field theories, string theory and gravity},'' {\em Phys. Rept.} {\bf 323}
  (2000) 183--386,
\href{http://www.arXiv.org/abs/hep-th/9905111}{{\tt hep-th/9905111}}.

\bibitem{Brown:1994gs}
J.~D. Brown, J.~Creighton, and R.~B. Mann, ``{Temperature, energy and heat
  capacity of asymptotically anti-de Sitter black holes},'' {\em Phys. Rev.}
  {\bf D50} (1994) 6394--6403,
\href{http://www.arXiv.org/abs/gr-qc/9405007}{{\tt gr-qc/9405007}}.

\bibitem{Olsson:2005en}
  M.~E.~Olsson, ``The stringy nature of the 2d type-0A black hole,''
\href{http://www.arXiv.org/abs/hep-th/0511106}{{\tt hep-th/0511106}}.

\bibitem{Harvey:1992xk}
J.~A. Harvey and A.~Strominger, ``Quantum aspects of black holes,'' in {\em
  Recent directions in particle theory: from superstrings and black holes to
  the standard model (TASI - 92)}.
\newblock 1992.
\newblock
\href{http://www.arXiv.org/abs/hep-th/9209055}{{\tt hep-th/9209055}}.
\newblock

A.~Strominger, ``Les {H}ouches lectures on black holes,''
  \href{http://www.arXiv.org/abs/arXiv:hep-th/9501071}{{\tt
  arXiv:hep-th/9501071}}.
Talk given at {NATO} Advanced Study Institute.

L.~Thorlacius, ``Black hole evolution,'' {\em Nucl. Phys. Proc. Suppl.} {\bf
  41} (1995) 245--275,
\href{http://www.arXiv.org/abs/hep-th/9411020}{{\tt hep-th/9411020}}.

S.~B. Giddings, ``Quantum mechanics of black holes,'' {\em Trieste HEP
  Cosmology} (1994) 0530--574,
\href{http://www.arXiv.org/abs/arXiv:hep-th/9412138}{{\tt
  arXiv:hep-th/9412138}}.

\bibitem{Duff:1993wm}
M.~J. Duff, ``Twenty years of the weyl anomaly,'' {\em Class. Quant. Grav.}
  {\bf 11} (1994) 1387--1404,
\href{http://www.arXiv.org/abs/hep-th/9308075}{{\tt hep-th/9308075}}.

\bibitem{Christensen:1977jc}
S.~M. Christensen and S.~A. Fulling, ``Trace anomalies and the {H}awking
  effect,'' {\em Phys. Rev.} {\bf D15} (1977)
2088--2104.

\bibitem{Mukhanov:1994ax}
V.~Mukhanov, A.~Wipf, and A.~Zelnikov, ``{On 4-D Hawking radiation from
  effective action},'' {\em Phys. Lett.} {\bf B332} (1994) 283--291,
\href{http://arXiv.org/abs/hep-th/9403018}{{\tt hep-th/9403018}}.

\bibitem{Chiba:1997ex}
T.~Chiba and M.~Siino, ``Disappearance of black hole criticality in
  semiclassical general relativity,'' {\em Mod. Phys. Lett.} {\bf A12} (1997)
709--718.

S.~Ichinose, ``{Weyl anomaly of 2D dilaton-scalar gravity and hermiticity of
  system operator},'' {\em Phys. Rev.} {\bf D57} (1998) 6224--6229,
\href{http://arXiv.org/abs/hep-th/9707025}{{\tt hep-th/9707025}}.

W.~Kummer, H.~Liebl, and D.~V. Vassilevich, ``Hawking radiation for
  non-minimally coupled matter from generalized 2d black hole models,'' {\em
  Mod. Phys. Lett.} {\bf A12} (1997) 2683--2690,
\href{http://arXiv.org/abs/hep-th/9707041}{{\tt hep-th/9707041}}.

\bibitem{Izawa:1999ib}
K.~I. Izawa, ``On nonlinear gauge theory from a deformation theory
  perspective,'' {\em Prog. Theor. Phys.} {\bf 103} (2000) 225--228,
\href{http://arXiv.org/abs/hep-th/9910133}{{\tt hep-th/9910133}}.

\bibitem{Lauscher:2001ya}
O.~Lauscher and M.~Reuter, ``Ultraviolet fixed point and generalized flow
  equation of quantum gravity,'' {\em Phys. Rev.} {\bf D65} (2002) 025013,
\href{http://www.arXiv.org/abs/hep-th/0108040}{{\tt hep-th/0108040}}.

\bibitem{Bonanno:2006eu}
A.~Bonanno and M.~Reuter, ``Spacetime structure of an evaporating black hole in
  quantum gravity,''
\href{http://www.arXiv.org/abs/hep-th/0602159}{{\tt hep-th/0602159}}.

\bibitem{Niedermaier:2003fz}
M.~Niedermaier, ``Dimensionally reduced gravity theories are asymptotically
  safe,'' {\em Nucl. Phys.} {\bf B673} (2003) 131--169,
\href{http://www.arXiv.org/abs/hep-th/0304117}{{\tt hep-th/0304117}}.

\bibitem{Frolov:1981mz}
V.~P. Frolov and G.~A. Vilkovisky, ``Spherically symmetric collapse in quantum
  gravity,'' {\em Phys. Lett.} {\bf B106} (1981)
307--313.

\bibitem{Parikh:1998ux}
M.~K. Parikh and F.~Wilczek, ``Global structure of evaporating black holes,''
  {\em Phys. Lett.} {\bf B449} (1999) 24--29,
\href{http://www.arXiv.org/abs/gr-qc/9807031}{{\tt gr-qc/9807031}}.

\bibitem{Ashtekar:2005cj}
A.~Ashtekar and M.~Bojowald, ``{Black hole evaporation: A paradigm},'' {\em
  Class. Quant. Grav.} {\bf 22} (2005) 3349--3362,
\href{http://www.arXiv.org/abs/gr-qc/0504029}{{\tt gr-qc/0504029}}.

\bibitem{Hayward:2005gi}
S.~A. Hayward, ``Formation and evaporation of regular black holes,'' {\em Phys.
  Rev. Lett.} {\bf 96} (2006) 031103,
\href{http://www.arXiv.org/abs/gr-qc/0506126}{{\tt gr-qc/0506126}}.

\bibitem{Carlip:2004mn}
S.~Carlip, ``Horizon constraints and black hole entropy,'' {\em Class. Quant.
  Grav.} {\bf 22} (2005) 1303--1312,
\href{http://www.arXiv.org/abs/hep-th/0408123}{{\tt hep-th/0408123}};
``Horizon constraints and black hole entropy,''
\href{http://www.arXiv.org/abs/gr-qc/0508071}{{\tt gr-qc/0508071}};
``Horizons, constraints, and black hole entropy,''
\href{http://www.arXiv.org/abs/gr-qc/0601041}{{\tt gr-qc/0601041}}.

\bibitem{Bergamin:2005pg}
L.~Bergamin, D.~Grumiller, W.~Kummer, and D.~V. Vassilevich, ``Physics-to-gauge
  conversion at black hole horizons,''
{\em Class. Quant. Grav.}  {\bf 23} (2006) 3075--3101,
\href{http://www.arXiv.org/abs/hep-th/0512230}{{\tt hep-th/0512230}}.

\bibitem{'tHooft:2004ek}
G.~'t~Hooft, ``Horizons,''
\href{http://www.arXiv.org/abs/gr-qc/0401027}{{\tt gr-qc/0401027}}.

\bibitem{Kuchar:1994zk}
K.~V. Kucha{\v{r}}, ``Geometrodynamics of {S}chwarzschild black holes,'' {\em
  Phys. Rev.} {\bf D50} (1994) 3961--3981,
\href{http://www.arXiv.org/abs/gr-qc/9403003}{{\tt gr-qc/9403003}}.

\bibitem{Choptuik:1993jv}
M.~W. Choptuik, ``Universality and scaling in gravitational collapse of a
  massless scalar field,'' {\em Phys. Rev. Lett.} {\bf 70} (1993)
9--12.

\bibitem{Gundlach:1998wm}
C.~Gundlach, ``Critical phenomena in gravitational collapse,'' {\em Adv. Theor.
  Math. Phys.} {\bf 2} (1998) 1--49,
\href{http://www.arXiv.org/abs/arXiv:gr-qc/9712084}{{\tt arXiv:gr-qc/9712084}}.

\bibitem{Strominger:1993tt}
A.~Strominger and L.~Thorlacius, ``Universality and scaling at the onset of
  quantum black hole formation,'' {\em Phys. Rev. Lett.} {\bf 72} (1994)
  1584--1587,
\href{http://www.arXiv.org/abs/hep-th/9312017}{{\tt hep-th/9312017}}.

\bibitem{Russo:1993yh}
J.~G. Russo, L.~Susskind, and L.~Thorlacius, ``Cosmic censorship in
  two-dimensional gravity,'' {\em Phys. Rev.} {\bf D47} (1993) 533--539,
\href{http://www.arXiv.org/abs/hep-th/9209012}{{\tt hep-th/9209012}}.

\bibitem{Peleg:1996ce}
Y.~Peleg, S.~Bose, and L.~Parker, ``{Choptuik scaling and quantum effects in 2D
  dilaton gravity},'' {\em Phys. Rev.} {\bf D55} (1997) 4525--4528,
\href{http://www.arXiv.org/abs/gr-qc/9608040}{{\tt gr-qc/9608040}}.

\bibitem{Birukou:2002ge}
M.~Birukou, V.~Husain, G.~Kunstatter, E.~Vaz, and M.~Olivier, ``Spherically
  symmetric scalar field collapse in any dimension,'' {\em Phys. Rev.} {\bf
  D65} (2002)
104036.

\bibitem{Sorkin:2005vz}
E.~Sorkin and Y.~Oren, ``{On Choptuik's scaling in higher dimensions},'' {\em
  Phys. Rev.} {\bf D71} (2005) 124005,
\href{http://www.arXiv.org/abs/hep-th/0502034}{{\tt hep-th/0502034}}.

\bibitem{Bland:2005kk}
J.~Bland, B.~Preston, M.~Becker, G.~Kunstatter, and V.~Husain, ``Dimension
  dependence of the critical exponent in spherically symmetric gravitational
  collapse,'' {\em Class. Quant. Grav.} {\bf 22} (2005)
5355--5364.

\bibitem{Husa:2000kr}
S.~Husa, C.~Lechner, M.~Purrer, J.~Thornburg, and P.~C. Aichelburg, ``{Type II
  critical collapse of a self-gravitating nonlinear sigma-model},'' {\em Phys.
  Rev.} {\bf D62} (2000) 104007,
\href{http://www.arXiv.org/abs/gr-qc/0002067}{{\tt gr-qc/0002067}}.

\bibitem{Purrer:2004nq}
M.~Purrer, S.~Husa, and P.~C. Aichelburg, ``{News from critical collapse: Bondi
  mass, tails and quasinormal modes},'' {\em Phys. Rev.} {\bf D71} (2005)
  104005,
\href{http://www.arXiv.org/abs/gr-qc/0411078}{{\tt gr-qc/0411078}}.

\bibitem{Kettner:2004aw}
J.~Kettner, G.~Kunstatter, and A.~J.~M. Medved, ``Quasinormal modes for single
  horizon black holes in generic 2-d dilaton gravity,'' {\em Class. Quant.
  Grav.} {\bf 21} (2004) 5317--5332,
\href{http://www.arXiv.org/abs/gr-qc/0408042}{{\tt gr-qc/0408042}}.

\bibitem{Nollert:1993}
H.-P. Nollert, ``{Quasinormal modes of Schwarzschild black holes: The
  determination of quasinormal frequencies with very large imaginary parts},''
  {\em Phys. Rev.} {\bf D47} (1993) 5253--5258.

N.~Andersson, ``{On the asymptotic distribution of quasinormal-mode frequencies
  for Schwarzschild black holes},'' {\em Class. Quant. Grav.} {\bf L10} (1993)
  61--67.

\bibitem{Motl:2002hd}
L.~Motl, ``{An analytical computation of asymptotic Schwarzschild quasinormal
  frequencies},'' {\em Adv. Theor. Math. Phys.} {\bf 6} (2003) 1135--1162,
\href{http://www.arXiv.org/abs/gr-qc/0212096}{{\tt gr-qc/0212096}}.

\bibitem{Motl:2003cd}
L.~Motl and A.~Neitzke, ``Asymptotic black hole quasinormal frequencies,'' {\em
  Adv. Theor. Math. Phys.} {\bf 7} (2003) 307--330,
\href{http://www.arXiv.org/abs/hep-th/0301173}{{\tt hep-th/0301173}}.

\bibitem{Unruh:1980cg}
W.~G. Unruh, ``Experimental black hole evaporation,'' {\em Phys. Rev. Lett.}
  {\bf 46} (1981)
1351--1353.

\bibitem{Novello:2002qg}
M.~Novello, M.~Visser, and G.~Volovik, eds., {\em Artificial black holes}.
\newblock World Scientific, River Edge, USA, 2002.

G.~E. Volovik, {\em The universe in a helium droplet}.
\newblock Clarendon, Oxford, UK, 2003.

C.~Barcelo, S.~Liberati, and M.~Visser, ``Analogue gravity,'' {\em Living Rev.
  Rel.} {\bf 8} (2005) 12,
\href{http://www.arXiv.org/abs/gr-qc/0505065}{{\tt gr-qc/0505065}}.

\bibitem{Fedichev:2003id}
P.~O. Fedichev and U.~R. Fischer, ``{Hawking radiation from sonic de Sitter
  horizons in expanding Bose-Einstein-condensed gases},'' {\em Phys. Rev.
  Lett.} {\bf 91} (2003) 240407,
\href{http://www.arXiv.org/abs/cond-mat/0304342}{{\tt cond-mat/0304342}}.

\bibitem{Balbinot:2004da}
R.~Balbinot, S.~Fagnocchi, A.~Fabbri, and G.~P. Procopio, ``Backreaction in
  acoustic black holes,'' {\em Phys. Rev. Lett.} {\bf 94} (2005) 161302,
\href{http://www.arXiv.org/abs/gr-qc/0405096}{{\tt gr-qc/0405096}}.

M.~Cadoni, ``Acoustic analogs of two-dimensional black holes,'' {\em Class.
  Quant. Grav.} {\bf 22} (2005) 409--420,
\href{http://www.arXiv.org/abs/gr-qc/0410138}{{\tt gr-qc/0410138}}.

M.~Cadoni and S.~Mignemi, ``Acoustic analogues of black hole singularities,''
  {\em Phys. Rev.} {\bf D72} (2005) 084012,
\href{http://www.arXiv.org/abs/gr-qc/0504143}{{\tt gr-qc/0504143}}.

\bibitem{Schutzhold:2005ex}
R.~Schutzhold, M.~Uhlmann, Y.~Xu, and U.~R. Fischer, ``{Quantum back-reaction
  in dilute Bose-Einstein condensates},'' {\em Phys. Rev.} {\bf D72} (2005)
  105005,
\href{http://www.arXiv.org/abs/cond-mat/0503581}{{\tt cond-mat/0503581}}.

\bibitem{Grumiller:Dresden}
D.~Grumiller, ``Black holes and analogues in two dimensions.'' talk presented
  at QUASIM05 in Dresden, July, 2005.

\bibitem{Kummer:1998zs}
W.~Kummer, H.~Liebl, and D.~V. Vassilevich, ``Integrating geometry in general
  2d dilaton gravity with matter,'' {\em Nucl. Phys.} {\bf B544} (1999)
  403--431,
\href{http://www.arXiv.org/abs/hep-th/9809168}{{\tt hep-th/9809168}}.

\bibitem{Grumiller:2000ah}
D.~Grumiller, W.~Kummer, and D.~V. Vassilevich, ``The virtual black hole in 2d
  quantum gravity,'' {\em Nucl. Phys.} {\bf B580} (2000) 438--456,
\href{http://www.arXiv.org/abs/gr-qc/0001038}{{\tt gr-qc/0001038}}.

\bibitem{Grumiller:2001rg}
D.~Grumiller, ``Virtual black hole phenomenology from 2d dilaton theories,''
  {\em Class. Quant. Grav.} {\bf 19} (2002) 997--1009,
\href{http://arXiv.org/abs/gr-qc/0111097}{{\tt gr-qc/0111097}};
``{Virtual Black Holes and the S-matrix},'' {\em Int. J. Mod.
  Phys.} {\bf D13} (2004) 1973--2002,
\href{http://www.arXiv.org/abs/hep-th/0409231}{{\tt hep-th/0409231}}.

\bibitem{Kummer:1997hy}
W.~Kummer, H.~Liebl, and D.~V. Vassilevich, ``Exact path integral quantization
  of generic 2-d dilaton gravity,'' {\em Nucl. Phys.} {\bf B493} (1997)
  491--502,
\href{http://arXiv.org/abs/gr-qc/9612012}{{\tt gr-qc/9612012}};
``Exact path integral quantization
  of 2-d dilaton gravity,''
\href{http://www.arXiv.org/abs/gr-qc/9710033}{{\tt gr-qc/9710033}}.

\bibitem{Kummer:2005tx}
W.~Kummer, ``Progress and problems in quantum gravity,''
\href{http://www.arXiv.org/abs/gr-qc/0512010}{{\tt gr-qc/0512010}}.

D.~Grumiller and W.~Kummer, ``How to approach quantum gravity: Background
  independence in 1+1 dimensions,'' in {\em What comes beyond the Standard
  Model? Symmetries beyond the standard model}, N.~M. Borstnik, H.~B. Nielsen,
  C.~D. Froggatt, and D.~Lukman, eds., vol.~4 of {\em Bled Workshops in
  Physics}, pp.~184--196, EURESCO.
\newblock Portoroz, Slovenia, July, 2003.
\newblock \href{http://www.arXiv.org/abs/gr-qc/0310068}{{\tt gr-qc/0310068}}.
\newblock
based upon two talks.

\bibitem{Kummer:2001ip}
W.~Kummer, ``On the gauge-independence of the {S}-matrix,'' {\em Eur. Phys. J.}
  {\bf C21} (2001) 175--179,
\href{http://arXiv.org/abs/hep-th/0104123}{{\tt hep-th/0104123}}.

\bibitem{Grumiller:2003mc}
D.~Grumiller, W.~Kummer, and D.~V. Vassilevich, ``Positive specific heat of the
  quantum corrected dilaton black hole,'' {\em JHEP} {\bf 07} (2003) 009,
\href{http://www.arXiv.org/abs/hep-th/0305036}{{\tt hep-th/0305036}}.

\bibitem{Fischer:2001vz}
P.~Fischer, D.~Grumiller, W.~Kummer, and D.~V. Vassilevich, ``S-matrix for
  s-wave gravitational scattering,'' {\em Phys. Lett.} {\bf B521} (2001)
  357--363, \href{http://arXiv.org/abs/gr-qc/0105034}{{\tt gr-qc/0105034}}.
Erratum ibid. {\bf B532} (2002) 373.

D.~Grumiller, {\em Quantum dilaton gravity in two dimensions with matter}.
\newblock PhD thesis, {T}echnische {U}niversit{\"a}t {W}ien, 2001.
\newblock
\href{http://www.arXiv.org/abs/gr-qc/0105078}{{\tt gr-qc/0105078}}.
\newblock

\bibitem{Balasin:1992nm}
H.~Balasin, W.~Kummer, O.~Piguet, and M.~Schweda, ``{On the regularization of
  the mass zero 2-D propagator},'' {\em Phys. Lett.} {\bf B287} (1992)
138--144.

\bibitem{Polyakov:1981rd}
A.~M. Polyakov, ``Quantum geometry of bosonic strings,'' {\em Phys. Lett.} {\bf
  B103} (1981)
207--210.


\bibitem{Nojiri:1992st}
S.~Nojiri and I.~Oda, ``Charged dilatonic black hole and Hawking radiation
in two- dimensions,'' {\em Phys. Lett.} {\bf B294} (1992) 317--324,
\href{http://arXiv.org/abs/hep-th/9206087}{{\tt hep-th/9206087}}.

A.~Ori, ``Evaporation of a two-dimensional charged black hole,'' {\em Phys. Rev.} {\bf D63} (2001) 104016,
\href{http://arXiv.org/abs/gr-qc/0102067}{{\tt gr-qc/0102067}}.

\bibitem{Wess:1971yu}
J.~Wess and B.~Zumino, ``{Consequences of anomalous Ward identities},'' {\em
  Phys. Lett.} {\bf B37} (1971)
95.

%

\bibitem{Coleman:1974bu}
  S.~R. Coleman,``Quantum Sine-Gordon Equation As The Massive Thirring Model,''
  {\em Phys. Rev.} {\bf D11} (1975) 2088.

 S. Mandelstam, ``Soliton Operators For The Quantized Sine-Gordon Equation,''
  {\em Phys. Rev.} {\bf D11} (1975) 3026.

\bibitem{Frolov:2005ps}
A.~V. Frolov, K.~R. Kristjansson, and L.~Thorlacius, ``Semi-classical geometry
  of charged black holes,'' {\em Phys. Rev.} {\bf D72} (2005) 021501,
\href{http://www.arXiv.org/abs/hep-th/0504073}{{\tt hep-th/0504073}}.

\bibitem{Guillemin:1974}
V.~Guillemin and A.~Pollack, {\em Differential Topology}.
\newblock Englewood Cliffs, NJ: Prentice-Hall, 1974.

\bibitem{Ambjorn:1998fd}
J.~Ambjorn, R.~Loll, J.~L. Nielsen, and J.~Rolf, ``{Euclidean and Lorentzian
  quantum gravity: Lessons from two dimensions},'' {\em Chaos Solitons
  Fractals} {\bf 10} (1999) 177--195,
\href{http://www.arXiv.org/abs/hep-th/9806241}{{\tt hep-th/9806241}}.

\bibitem{Gastmans:1977ad}
R.~Gastmans, R.~Kallosh, and C.~Truffin, ``Quantum gravity near
  two-dimensions,'' {\em Nucl. Phys.} {\bf B133} (1978)
417.

S.~M. Christensen and M.~J. Duff, ``Quantum gravity in two + epsilon
  dimensions,'' {\em Phys. Lett.} {\bf B79} (1978)
213.

S.~Weinberg in {\em General Relativity, an Einstein Centenary Survey},
  S.~Hawking and W.~Israel, eds.
\newblock Cambridge University Press, 1979.

\bibitem{Mann:1992ar}
R.~B. Mann and S.~F. Ross, ``{The D $\to$ 2 limit of general relativity},''
  {\em Class. Quant. Grav.} {\bf 10} (1993) 345--351,
\href{http://www.arXiv.org/abs/gr-qc/9208004}{{\tt gr-qc/9208004}}.

\bibitem{Jackiw:2005su}
R.~Jackiw, ``{Weyl symmetry and the Liouville theory},''
\href{http://www.arXiv.org/abs/hep-th/0511065}{{\tt hep-th/0511065}}.

\bibitem{Deser:1982wh}
S.~Deser, R.~Jackiw, and S.~Templeton, ``Topologically massive gauge
  theories,'' {\em Ann. Phys.} {\bf 140} (1982)
372--411;
{\em Erratum-ibid.} {\bf 185} (1988) 406;
``Three-dimensional massive gauge
  theories,'' {\em Phys. Rev. Lett.} {\bf 48} (1982)
975--978.

\bibitem{Banados:1992wn}
M.~Banados, C.~Teitelboim, and J.~Zanelli, ``The black hole in
  three-dimensional space-time,'' {\em Phys. Rev. Lett.} {\bf 69} (1992)
  1849--1851,
\href{http://www.arXiv.org/abs/hep-th/9204099}{{\tt hep-th/9204099}}.

M.~Banados, M.~Henneaux, C.~Teitelboim, and J.~Zanelli, ``Geometry of the (2+1)
  black hole,'' {\em Phys. Rev.} {\bf D48} (1993) 1506--1525,
\href{http://www.arXiv.org/abs/gr-qc/9302012}{{\tt gr-qc/9302012}}.

\bibitem{Sahoo:2006vz}
B.~Sahoo and A.~Sen, ``{BTZ black hole with Chern-Simons and higher derivative
  terms},''
\href{http://www.arXiv.org/abs/hep-th/0601228}{{\tt hep-th/0601228}}.

\bibitem{Kummer:1992rt}
W.~Kummer and D.~J. Schwarz, ``Renormalization of {R}**2 gravity with dynamical
  torsion in d = 2,'' {\em Nucl. Phys.} {\bf B382} (1992)
171--186;
``{General analytic solution of R**2 gravity with
  dynamical torsion in two-dimensions},'' {\em Phys. Rev.} {\bf D45} (1992)
3628--3635.

\bibitem{Kotov:2004wz}
A.~Kotov, P.~Schaller, and T.~Strobl, ``Dirac sigma models,'' {\em Commun.
  Math. Phys.} {\bf 260} (2005) 455--480,
\href{http://www.arXiv.org/abs/hep-th/0411112}{{\tt hep-th/0411112}}.

\bibitem{Nicolai:1996pd}
H.~Nicolai, D.~Korotkin, and H.~Samtleben, ``Integrable classical and quantum
  gravity,''
\href{http://arXiv.org/abs/hep-th/9612065}{{\tt hep-th/9612065}}.

D.~Korotkin and H.~Samtleben, ``Canonical quantization of cylindrical
  gravitational waves with two polarizations,'' {\em Phys. Rev. Lett.} {\bf 80}
  (1998) 14--17,
\href{http://arXiv.org/abs/gr-qc/9705013}{{\tt gr-qc/9705013}};
``Yangian symmetry in integrable quantum
  gravity,'' {\em Nucl. Phys.} {\bf B527} (1998) 657--689,
\href{http://arXiv.org/abs/hep-th/9710210}{{\tt hep-th/9710210}}.

D.~Bernard and N.~Regnault, ``Vertex operator solutions of 2-d dimensionally
  reduced gravity,'' {\em Commun. Math. Phys.} {\bf 210} (2000) 177--201,
\href{http://arXiv.org/abs/solv-int/9902017}{{\tt solv-int/9902017}}.

L.~D. Faddeev, R.~M. Kashaev, and A.~Y. Volkov, ``{Strongly coupled quantum
  discrete Liouville theory. I: Algebraic approach and duality},'' {\em Commun.
  Math. Phys.} {\bf 219} (2001) 199--219,
\href{http://arXiv.org/abs/hep-th/0006156}{{\tt hep-th/0006156}}.

J.~Teschner, ``Liouville theory revisited,'' {\em Class. Quant. Grav.} {\bf 18}
  (2001) R153--R222,
\href{http://arXiv.org/abs/hep-th/0104158}{{\tt hep-th/0104158}}.

\bibitem{Gurses:2005yg}
M.~Gurses and S.~Tek, ``{KdV Surfaces},''
\href{http://www.arXiv.org/abs/nlin.si/0511049}{{\tt nlin.si/0511049}}.

\bibitem{Schmidt:1999wb}
H.-J. Schmidt, ``The classical solutions of two-dimensional gravity,'' {\em
  Gen. Rel. Grav.} {\bf 31} (1999) 1187--1210,
\href{http://arXiv.org/abs/gr-qc/9905051}{{\tt gr-qc/9905051}}.

\bibitem{Obukhov:1997uc}
Y.~N. Obukhov and F.~W. Hehl, ``Black holes in two dimensions,''
\href{http://arXiv.org/abs/hep-th/9807101}{{\tt hep-th/9807101}}.

\bibitem{Dereli:1994vx}
T.~Dereli and R.~W. Tucker, ``Nonmetricity induced by dilaton gravity in
  two-dimensions,'' {\em Class. Quant. Grav.} {\bf 11} (1994)
2575--2583.

Y.~N. Obukhov, ``Two-dimensional metric-affine gravity,'' {\em Phys. Rev.} {\bf
  D69} (2004) 064009,
\href{http://www.arXiv.org/abs/gr-qc/0311091}{{\tt gr-qc/0311091}}.

M.~Adak, ``Nonmetricity and torsion induced by dilaton gravity in two
  dimension,''
\href{http://www.arXiv.org/abs/gr-qc/0509010}{{\tt gr-qc/0509010}}.

\bibitem{Douglas:2001ba}
M.~R. Douglas and N.~A. Nekrasov, ``Noncommutative field theory,'' {\em Rev.
  Mod. Phys.} {\bf 73} (2001) 977--1029,
\href{http://www.arXiv.org/abs/hep-th/0106048}{{\tt hep-th/0106048}}.

R.~J. Szabo, ``Quantum field theory on noncommutative spaces,'' {\em Phys.
  Rept.} {\bf 378} (2003) 207--299,
\href{http://www.arXiv.org/abs/hep-th/0109162}{{\tt hep-th/0109162}}.

\bibitem{Buric:2004rm}
M.~Buric and J.~Madore, ``Noncommutative 2-dimensional models of gravity,''
\href{http://www.arXiv.org/abs/hep-th/0406232}{{\tt hep-th/0406232}};
``A dynamical 2-dimensional fuzzy space,'' {\em Phys.
  Lett.} {\bf B622} (2005) 183--191,
\href{http://www.arXiv.org/abs/hep-th/0507064}{{\tt hep-th/0507064}}.

\bibitem{Cacciatori:2002ib}
S.~Cacciatori {\em et al.}, ``Noncommutative gravity in two dimensions,'' {\em
  Class. Quant. Grav.} {\bf 19} (2002) 4029--4042,
\href{http://www.arXiv.org/abs/hep-th/0203038}{{\tt hep-th/0203038}}.

\bibitem{Vassilevich:2004ym}
D.~V. Vassilevich, ``Quantum noncommutative gravity in two dimensions,'' {\em
  Nucl. Phys.} {\bf B715} (2005) 695--712,
\href{http://www.arXiv.org/abs/hep-th/0406163}{{\tt hep-th/0406163}}.

\bibitem{Vassilevich:2005fk}
D.~V. Vassilevich, ``Constraints, gauge symmetries, and noncommutative gravity
  in two dimensions,''
\href{http://www.arXiv.org/abs/hep-th/0502120}{{\tt hep-th/0502120}}.

\bibitem{Vassilevich:2006uv}
D.~V. Vassilevich, R.~Fresneda, and D.~M. Gitman, ``{Stability of a
  noncommutative Jackiw-Teitelboim gravity},''
\href{http://www.arXiv.org/abs/hep-th/0602095}{{\tt hep-th/0602095}}.

\bibitem{Grumiller:2003df}
D.~Grumiller, W.~Kummer, and D.~V. Vassilevich, ``A note on the triviality of
  kappa-deformations of gravity,'' {\em Ukr. J. Phys.} {\bf 48} (2003)
  329--333,
\href{http://www.arXiv.org/abs/hep-th/0301061}{{\tt hep-th/0301061}}.

\bibitem{Aschieri:2005yw}
P.~Aschieri {\em et al.}, ``A gravity theory on noncommutative spaces,'' {\em
  Class. Quant. Grav.} {\bf 22} (2005) 3511--3532,
\href{http://www.arXiv.org/abs/hep-th/0504183}{{\tt hep-th/0504183}}.

\bibitem{Zupnik:2005ph}
B.~M. Zupnik, ``Reality in noncommutative gravity,''
\href{http://www.arXiv.org/abs/hep-th/0512231}{{\tt hep-th/0512231}}.

\bibitem{Chaichian:2004za}
M.~Chaichian, P.~P. Kulish, K.~Nishijima, and A.~Tureanu, ``{On a
  Lorentz-invariant interpretation of noncommutative space-time and its
  implications on noncommutative QFT},'' {\em Phys. Lett.} {\bf B604} (2004)
  98--102,
\href{http://www.arXiv.org/abs/hep-th/0408069}{{\tt hep-th/0408069}}.

J.~Wess, ``{Deformed coordinate spaces: Derivatives},''
\href{http://www.arXiv.org/abs/hep-th/0408080}{{\tt hep-th/0408080}}.

\bibitem{Vassilevich:2006tc}
D.~V. Vassilevich, ``Twist to close,''
\href{http://www.arXiv.org/abs/hep-th/0602185}{{\tt hep-th/0602185}}.

P.~Aschieri, M.~Dimitrijevic, F.~Meyer, S.~Schraml, and J.~Wess, ``Twisted
  gauge theories,''
\href{http://www.arXiv.org/abs/hep-th/0603024}{{\tt hep-th/0603024}}.

\bibitem{Chaichian:2004yh}
M.~Chaichian, P.~Presnajder, and A.~Tureanu, ``{New concept of relativistic
  invariance in NC space-time: Twisted Poincare symmetry and its
  implications},'' {\em Phys. Rev. Lett.} {\bf 94} (2005) 151602,
\href{http://www.arXiv.org/abs/hep-th/0409096}{{\tt hep-th/0409096}}.

P.~Aschieri, M.~Dimitrijevic, F.~Meyer, and J.~Wess, ``Noncommutative geometry
  and gravity,'' {\em Class. Quant. Grav.} {\bf 23} (2006) 1883--1912,
\href{http://www.arXiv.org/abs/hep-th/0510059}{{\tt hep-th/0510059}}.

  A.~Kobakhidze, ``Theta-twisted gravity,''
\href{http://www.arXiv.org/abs/hep-th/0603132}{{\tt hep-th/0603132}}.

\bibitem{Oeckl:2000eg}
R.~Oeckl, ``{Untwisting noncommutative R**d and the equivalence of quantum
  field theories},'' {\em Nucl. Phys.} {\bf B581} (2000) 559--574,
\href{http://www.arXiv.org/abs/hep-th/0003018}{{\tt hep-th/0003018}}.

\bibitem{Bojowald:2004wu}
M.~Bojowald, A.~Kotov, and T.~Strobl, ``{Lie algebroid morphisms, Poisson Sigma
  Models, and off- shell closed gauge symmetries},'' {\em J. Geom. Phys.} {\bf
  54} (2005) 400--426,
\href{http://www.arXiv.org/abs/math.dg/0406445}{{\tt math.dg/0406445}}.

\end{thebibliography}

\providecommand{\href}[2]{#2}\begingroup\raggedright\endgroup

\end{document}